\documentclass[journal,twoside,web]{IEEEtran}
\usepackage{cite}
\usepackage{amsmath,amssymb,amsfonts}
\usepackage{algorithmic}
\usepackage{graphicx}
\usepackage{textcomp}
\usepackage{hyperref}
\usepackage{multirow}
\usepackage{url}
\usepackage{underscore}

\def\BibTeX{{\rm B\kern-.05em{\sc i\kern-.025em b}\kern-.08em
    T\kern-.1667em\lower.7ex\hbox{E}\kern-.125emX}}

\begin{document}
\title{Benchmark of Segmentation Techniques for Pelvic Fracture in CT and X-Ray: Summary of the PENGWIN 2024 Challenge}

\author{Yudi Sang, Yanzhen Liu, Sutuke Yibulayimu, Yunning Wang, Benjamin D. Killeen, Mingxu Liu, Ping-Cheng Ku, Ole Johannsen, Karol Gotkowski, Maximilian Zenk, Klaus Maier-Hein, Fabian Isensee, Peiyan Yue, Yi Wang, Haidong Yu, Zhaohong Pan, Yutong He, Xiaokun Liang, Daiqi Liu, Fuxin Fan, Artur Jurgas, Andrzej Skalski, Yuxi Ma, Jing Yang, Szymon Płotka, Rafał Litka, Gang Zhu, Yingchun Song, Mathias Unberath, Mehran Armand, Dan Ruan, S. Kevin Zhou, Qiyong Cao, Chunpeng Zhao, Xinbao Wu, and Yu Wang
\thanks{This work was supported by Nation Natural Science Foundation of China (NSFC6247010104), National Key Research and Development Program of China (2022YFC2504304), Natural Science Foundation of Beijing (L222136), and Beijing Science and Technology Project (Z221100003522007 and Z241100009024030). \textit{Corresponding author: Yu Wang, email: wangyu@buaa.edu.cn}.}
\thanks{Y. Sang, G. Zhu, and Y. Song are with Beijing Rossum Robot Technology Co., Ltd., Beijing, China.
Y. Liu, S. Yibulayimu, Y. Wang, and Y. Wang are with the Key Laboratory of Biomechanics and Mechanobiology, Ministry of Education, Beijing Advanced Innovation Center for Biomedical Engineering, School of Biological Science and Medical Engineering, Beihang University, Beijing, China.
B.~D. Killeen, M. Liu, P.-C. Ku, M. Unberath, and M. Armand are with the Department of Computer Science, Johns Hopkins University, Baltimore, MD, USA.
O. Johannsen, K. Gotkowski, M. Zenk, and K. Maier-Hein are with the Division of Medical Image Computing, German Cancer Research Center (DKFZ), Heidelberg, Germany. O. Johannsen, K. Gotkowski, and F. Isensee are also with Helmholtz Imaging, Heidelberg, Germany.
P. Yue, Y. Wang, and H. Yu are with the Smart Medical Imaging, Learning and Engineering (SMILE) Lab, Medical UltraSound Image Computing (MUSIC) Lab, School of Biomedical Engineering, Shenzhen University Medical School, Shenzhen University, Shenzhen, China.
Z. Pan, Y. He, and X. Liang are with the Institute of Biomedical and Health Engineering, Shenzhen Institute of Advanced Technology, Chinese Academy of Sciences, Shenzhen, China.
D. Liu and F. Fan are with the Pattern Recognition Lab, Friedrich-Alexander-University Erlangen-Nuremberg, Erlangen, Germany.
A. Jurgas and A. Skalski are with the Department of Measurement and Electronics, AGH University of Krakow, Krakow, Poland, and also with MedApp S.A., Krakow, Poland. 
Y. Ma and J. Yang are with the National Institute for Data Science in Health and Medicine, Xiamen University, Xiamen, China.
S. Płotka is with the Faculty of Mathematics and Computer Science, Jagiellonian University, Krakow, Poland.
R. Litka is with the Faculty of Electronics and Information Technology, Warsaw University of Technology, Warsaw, Poland, and the Sano Centre for Computational Medicine, Krakow, Poland
D. Ruan is with the Department of Radiation Oncology, University of California, Los Angeles, CA, USA.
S. K. Zhou is with the School of Biomedical Engineering, Division of Life Sciences and Medicine, University of Science and Technology of China, Hefei, China.
Q. Cao, C. Zhao, and X. Wu are with the Department of Orthopaedics and Traumatology, Beijing Jishuitan Hospital, Beijing, China.}}

\maketitle

\begin{abstract}
The segmentation of pelvic fracture fragments in CT and X-ray images is crucial for trauma diagnosis, surgical planning, and intraoperative guidance. However, accurately and efficiently delineating the bone fragments remains a significant challenge due to complex anatomy and imaging limitations. The PENGWIN challenge, organized as a MICCAI 2024 satellite event, aimed to advance automated fracture segmentation by benchmarking state-of-the-art algorithms on these complex tasks. A diverse dataset of 150 CT scans was collected from multiple clinical centers, and a large set of simulated X-ray images was generated using the DeepDRR method. Final submissions from 16 teams worldwide were evaluated under a rigorous multi-metric testing scheme. The top-performing CT algorithm achieved an average fragment-wise intersection over union (IoU) of 0.930, demonstrating satisfactory accuracy. However, in the X-ray task, the best algorithm achieved an IoU of 0.774, which is promising but not yet sufficient for intra-operative decision-making, reflecting the inherent challenges of fragment overlap in projection imaging. Beyond the quantitative evaluation, the challenge revealed methodological diversity in algorithm design. Variations in instance representation, such as primary-secondary classification versus boundary-core separation, led to differing segmentation strategies. Despite promising results, the challenge also exposed inherent uncertainties in fragment definition, particularly in cases of incomplete fractures. These findings suggest that interactive segmentation approaches, integrating human decision-making with task-relevant information, may be essential for improving model reliability and clinical applicability. 
\end{abstract}


\section{Introduction}
\label{sec:introduction}
\IEEEPARstart{{P}}{elvic} fractures are among the most severe orthopedic injuries, often resulting from high-energy trauma such as motor vehicle accidents and falls from height. These fractures are associated with a disability rate exceeding 50\% and a mortality rate above 13\%, making them the deadliest of all compound fractures~\cite{kowal2007basics, sugano2013computer}. The complex anatomy of the pelvis, combined with surrounding soft tissues, makes surgical intervention particularly challenging. In recent years, robotic-assisted closed fracture reduction surgeries have demonstrated improved clinical outcomes with reduced bleeding, lower infection rates, decreased vascular and nerve damage, and faster postoperative recovery~\cite{LSBT2022robot}. The precision and outcome of these image-guided robotic surgery heavily rely on accurate preoperative and intraoperative imaging and segmentation.

Automated pelvic fracture segmentation is a crucial step in trauma diagnosis and surgical planning. 
In 3D CT scans, precise fracture segmentation is a prerequisite for fragment-wise reduction planning, enabling accurate pose and trajectory planning, as well as screw planning, for surgical realignment and fixation~\cite{liu2025preoperative, shi2025automatic}. It also plays a key role in detailed fracture classification, aiding in clinical decision-making.
In 2D C-arm X-ray imaging, fracture segmentation is equally important for diagnostic interpretation and intra-operative guidance. By segmenting fractures in radiological images, surgical plans can be accurately transferred to the operating space, facilitating C-arm imaging-based navigation and registration~\cite{han2021fracture}. 

Traditionally, fracture segmentation in CT has been performed semi-automatically using 3D interactive tools. In the current surgical planning workflow for robotic pelvic fracture reduction, the fractured pelvis is typically extracted using a graph-cut algorithm, with manually selected foreground and background seed points. Following this step, each bone fragment must be individually isolated. For completely separated fragments, connected component analysis can be used for extraction. However, when fragments remain connected or collided, operators must carefully delineate the fracture surfaces, either by manually tracing contours slice by slice in 2D images or by examining the fracture structure in a 3D view. This process is highly labor-intensive, often requiring 0.5 to 1 hour for complex fractures with overlapping surfaces, not to mention the additional time required for subsequent fragment pose planning. The efficiency of this workflow is particularly critical, as pelvic fracture reductions are often non-elective procedures, sometimes required in emergency settings. Moreover, the segmentation accuracy is heavily dependent on the skill and experience of the operator.

For intraoperative X-ray imaging, fracture segmentation holds significant potential to facilitate pelvic reduction surgery; however, it has not yet been adopted into clinical routine. Pelvic fracture fixation is a highly complex procedure that relies on radiographic imaging to accurately place Kirschner wires and screws within specific bony corridors, following a detailed preoperative plan. In intraoperative cone-beam CT (CBCT) imaging, efforts have been successfully made to automatically extract bone regions and register them with preoperative CT scans~\cite{liu2024automatic}. However, similar techniques for X-ray remain underexplored. While C-arm X-ray imaging is far more widely used than CBCT, its segmentation process presents several challenges, including the difficulty in distinguishing overlapping structures, the limited visibility of fracture lines, and the need for precise alignment of fractured fragments. These challenges highlight the need for improved segmentation techniques in X-ray to enhance intraoperative guidance with lower radiation exposure and wider applicability.

Automating pelvic fracture segmentation remains an unsolved challenge due to the diverse shapes and positions of bone fragments and the complex fracture surfaces caused by bone collisions. Despite the success of deep learning in various medical imaging applications, its use in automated pelvic fracture segmentation remains relatively limited. This is primarily attributed to the lack of extensive image datasets and annotations specific to fractured pelvises. 

The \textbf{Pe}lvic Bo\textbf{n}e Fra\textbf{g}ments \textbf{w}ith \textbf{In}juries (PENGWIN) Challenge 2024 was launched to address these challenges by fostering the development of robust, automated segmentation techniques for pelvic fractures in both 3D CT and 2D X-ray modalities. Hosted as a satellite event at the 27th International Conference on Medical Image Computing and Computer Assisted Intervention (MICCAI 2024), this challenge aimed to: (1) advance the state of the art in pelvic fracture segmentation using deep learning; (2) provide a high-quality, publicly available dataset to facilitate further research in medical image analysis and orthopedics; and (3) promote interdisciplinary collaboration between researchers in medical imaging, artificial intelligence, and orthopedic surgery.

To achieve these goals, we curated a comprehensive, multi-institutional dataset consisting of CT scans from 150 patients scheduled for pelvic reduction surgery. 
These scans were acquired from multiple research centers using various scanning devices to ensure diversity in imaging conditions and patient cohorts. 
The ground-truth segmentations of the hipbones, sacrum, and fracture fragments were semi-automatically annotated and validated by medical experts. In addition, we generated 48,600 high-quality, realistic X-ray images using the DeepDRR method~\cite{unberath2018deepdrr} , simulating diverse virtual C-arm angles and incorporating surgical tool interference to closely reflect real-world intraoperative conditions.
Participants were challenged to develop fully automated segmentation algorithms for 3D CT and 2D X-ray tasks, contributing novel methods that improve the accuracy, robustness, and clinical applicability of fracture segmentation. 

In this paper, we present an in-depth review of the challenge, providing a comprehensive summary of related challenges and studies, participation statistics, algorithmic approaches, comparative performance analysis, and key insights gained.

\section{Related Works}

\subsection{Challenges on Orthopedic Fracture}
\label{sec:challenges}

We surveyed algorithmic challenges related to orthopedic image processing between 2017 and 2024. Challenges directly related to bone fractures are reviewed in detail.

The RibFrac Challenge~\cite{yang2024deep} was launched with MICCAI 2020, aimed to advance automated rib fracture detection and diagnosis. The dataset comprises over 5,000 rib fractures across 660 CT scans. The challenge included two tasks. In the first task, fracture detection was formulated as an instance segmentation problem, with annotated fractured regions provided as ground truth. Due to the inherent ambiguity in delineating fracture boundaries, the segmentation outputs were primarily used to evaluate detection performance, with FROC analysis employed for assessment. In the second task, the detected rib fractures were further classified into four clinically categories: buckle fractures, non-displaced fractures, displaced fractures, and segmental fractures.

The 2022 RSNA Cervical Spine Fracture AI Challenge~\cite{hu2024assessing} focused on rapid detection and localization of cervical spine fractures, which is critical to prevent neurological deterioration and paralysis following trauma. The dataset includes over 3,000 CT scans, with an approximately equal distribution of positive and negative cases. Participants were tasked with predicting the presence or absence of fractures at each of the seven cervical vertebrae, but were not required to locate them within the CT volume. 

The Fractured Bone Detection Challenge~\cite{fractured-bone-detection-challenge} was a lightweight, independent challenge launched on Kaggle. It provided a small dataset of 24 CT scans covering both upper and lower limbs. The objective was to identify fractures in 2D CT slices to facilitate rapid diagnosis. Each CT slice was binary classified based on the presence of fractures, with the area under curve (AUC) used as the evaluation metric.

In summary, challenges focusing on fractures remain limited in number, with no prior challenge specifically targeting pelvic bone or X-ray segmentation. Existing efforts have predominantly focused on fracture detection at a coarse level. However, the emerging adoption of image-guided reduction surgery highlights the growing demand for voxel-wise fracture fragment segmentation, calling for new challenges and benchmarks to drive advancements in this domain.

\subsection{Automated Fracture Segmentation}

\subsubsection{Fracture Segmentation in CT}
\label{sec:related_CT}

Traditional non-learning-based methods for automated and semi-automated CT fracture segmentation have been comprehensively summarized in previous studies~\cite{liu2025preoperative}. These approaches include intensity thresholding, graph cut, and region growing methods. While such methods can be effective in cases involving non-contact fractures, they often struggle in more complex scenarios, such as fractures with fragment contact, impaction, or compression~\cite{2011fullyMaxFlow}. In contrast, deep learning-based fracture segmentation remains a relatively underexplored area; however, recent studies have demonstrated its significant potential. In the following, we review the latest advancements in learning-based approaches for automated fracture segmentation in CT.

Early studies have explored the direct application of established segmentation networks to fracture fragment segmentation in CT scans. 
Yang \textit{et al.} \cite{yang2022recognition} applied a Mask R-CNN-based instance segmentation model to locate and extract individual fragments from 2D CT slices of intertrochanteric fractures.
Kim \textit{et al.} \cite{kim2023automatic} utilized DeepLab v3+ for semantic segmentation of tibial and fibular fragments in 2D CT slices. In their approach, fragment labels were manually assigned during data preparation; however, this strategy raises concerns regarding robustness due to potential label inconsistencies across patients and the limited scalability of treating individual fragments as semantic classes.
Wang \textit{et al.} \cite{wang2022accuracy} adopted a 3D V-Net model to segment intertrochanteric femoral fractures, using a three-class labeling scheme to distinguish the proximal femur, fracture fragment, and distal femur.

In our previous study, FracSegNet, fracture segmentation is performed using a two-stage pipeline comprising two sequential networks~\cite{liu2023pelvic}. First, an anatomical segmentation network extracts the hipbones and sacrum from CT images. Subsequently, a fracture segmentation network differentiates the primary fragment and secondary fragments within each bone region. The fracture segmentation network is trained using a distance-weighted loss, which emphasizes attention near fracture boundaries and improves performance in these critical regions. This approach enables semantic segmentation of major and minor fragments, thereby avoiding the need to explicitly handle a variable number of fragments during network inference. Following segmentation, secondary fragments are further separated using connected component analysis. However, this post-processing step can be insufficient when secondary fragments are in contact, leading to failures in accurately separating small fragments.

Building upon the same primary-secondary representation, Zeng \textit{et al.} \cite{zeng2024fragment} proposed a pelvic fracture segmentation network that incorporated the prediction of a fragment prior distance map as an auxiliary task to better capture the spatial relationships between fragments. In addition, a dynamic multi-scale feature fusion module was introduced to address the challenge posed by the varying sizes of fragments.

In addition, Deng \textit{et al.} \cite{deng2023synergistically} explored the synergistic integration of fragment segmentation and reduction pose planning for femoral fractures. Their method takes as input segmented bone surface point clouds from both the fractured and contralateral healthy sides. By using the healthy side as a template, the segmentation of fracture fragments is refined through an alternating process that minimizes the distance between the current fracture model and the template. However, this approach is only applicable when a complete contralateral template is available and may be limited by natural anatomical asymmetry.

\subsubsection{Fracture Segmentation in X-Ray}

Early applications of deep learning for fracture analysis in X-ray primarily focused on fracture detection and classification, with some approaches incorporating bone segmentation as a pre-processing step~\cite{joshi2020survey}.

Fracture detection can be performed at the image level, where a model determines the presence or absence of fractures. For instance, Cheng \textit{et al.} \cite{cheng2019application} developed a convolutional network to detect hip fractures from plain frontal pelvic radiographs.
To enhance interpretability, they further applied gradient-weighted class activation mapping (Grad-CAM) to visualize fracture regions.
Similar binary fracture classification approaches have been studied on other anatomies like long bones and spine~\cite{wang2022osteoporotic}.

Beyond image-level classification, fractures can be localized more precisely using bounding boxes or segmentation contours. This is particularly valuable for hairline or incomplete fractures, where the fracture line is clearly visible in the radiograph, though the entire fragment cannot be isolated. Traditional image processing and machine learning-based methods have been explored for fracture detection using bounding box representations~\cite{myint2018analysis,hrvzic2019local,hardalacc2022fracture}.
For example, Joshi \textit{et al.} \cite{joshi2022deep} employed a feature pyramid network to perform both fracture detection (bounding boxes) and segmentation (contours) in wrist bone fractures. Turk \textit{et al.} \cite{turk2024impact} investigated the effect of different backbone architectures on fracture segmentation in X-rays. Their study utilized 4,083 images covering hand, leg, hip, and shoulder fractures.

Additionally, in the pelvic region, a recent study proposed the use of deep networks to segment the entire pelvis in X-ray in the presence of fractures, demonstrating the potential of deep learning methods in handling fracture cases~\cite{lee2024deep}.
In spinal X-ray imaging, efforts have also been made to detect, segment, and assess vertebral compression fractures. Since spinal fractures are often non-displaced and lack clearly visible fracture lines, existing methods have primarily focused on extracting the full outline of the affected vertebrae~\cite{kim2021automatic}.

In short, similar to the related challenges discussed in Sec.~\ref{sec:challenges}, existing studies on X-ray fracture analysis have prioritized fracture detection, with some leveraging deep learning-based segmentation as a means to refine localization. While segmentation provides a more precise localization than bounding boxes to support trauma diagnosis, current methods lack the ability to delineate fracture lines accurately or extract individual fractured fragments.

\section{Challenge Description}

\subsection{Challenge Setup}
The PENGWIN Challenge 2024 is a one-time event with fixed submission deadline comprising two tasks, with participants free to compete in either or both. Each task included a development phase and a final test phase. The training data was released on April 5, 2024, enabling participants to develop and refine their models. The validation phase, beginning on May 15, 2024, allowed teams to test their models on a separate validation set multiple times, with the results displayed on a public leaderboard. In the final test phase, participants submitted their models as Docker containers on the Grand Challenge platform\footnote{\href{https://pengwin.grand-challenge.org/}{Challenge website: {https://pengwin.grand-challenge.org/}}.} by August 19, 2024. The final results were announced on August 31, 2024, and the top three teams in each task were awarded cash prizes and certificates, conditional upon the open-sourcing of their code in a public repository. The top three teams were invited to present their methods at MICCAI 2024 in October. 
Only the organizing team have access to the test cases. Members of the organizers' institutes could participate but were not eligible for awards. 

In the CT task, participants were provided with 3D pelvic CT scans containing fractures. The expected output was a voxel-wise segmentation mask, labeling the sacrum, hipbones, and individual bone fragments within the scan.

In the X-ray task, participants received 2D simulated X-ray images derived from CT scans. The objective was to generate a pixel-wise segmentation mask, identifying the sacrum, hipbones, and bone fragments in each X-ray image. Note that in this separate task, participants did not have access to the corresponding CT image for running the X-ray algorithms. 

Participants were permitted to use publicly available datasets and pre-trained network models, provided these were under a permissive license. Templates for preparing a container submission and evaluation scripts were provided to participants before the validation phase. The online platform allowed, at the maximum, a runtime environment with a NVIDIA T4 Tensor Core GPU, eight vCPUs, 32 GB of memory, and 225 GB NVMe SSD for running the container. The time limit was set to 10 minutes per case, including the container loading, image loading, and result saving times.

\subsection{Data}

\subsubsection{Curation of the CT Dataset}

The CT data were collected from 150 patients who were planning to undergo pelvic reduction surgery between 2017 and 2023 from six medical centers in China, with varying distributions (ranging from 6 to 58 scans per site) due to differences in data availability and institutional capacity. 
The CT scans were acquired from a diverse array of machines symmarized in Table~\ref{tab:CT_scanner}. The retrospective data usage has been approved by the Ethics Committee of the institutes. 

\begin{table}[t]
    \caption{List of CT scanners and number of cases in different centers.}\label{tab:CT_scanner}
    \centering
    \begin{tabular}{lll}
        \hline
        Center & CT Scanner & Number \\  
        \hline
        \multirow{2}{*}{JST} & TOSHIBA Aquilion PRIME & 58 \\  
                             & UNITED IMAGING uCT 550 & 45 \\  
        FSHTCM & PHILIPS Brilliance 64 & 25 \\  
        JLUFH & SIEMENS Sensation 64 & 7 \\  
        JLUTH & TOSHIBA Aquilion ONE & 7 \\  
        NFH & SIEMENS SOMATOM Force & 1 \\  
        TJH & GE Optima CT660 & 7 \\  
        \hline
    \end{tabular}
\end{table}

Clinically used routine CT protocols have been used in each CT scanner. All scans were acquired in high quality before fracture reduction surgery and do not present significant artifacts. During the acquisition process, all patients were positioned in supine pose. 

The dataset includes patients aged 16 to 94 (44.1 ± 16.8) years, with a gender distribution of 63 females and 87 males. The dataset encompasses a variety of fracture types, including pelvic ring dislocation (PRD, \textit{i.e.}, dislocation of pubic symphysis and/or sacroiliac joint without fracture within single bone, 5 cases), unilateral hip fracture (UHF, 54 cases), bilateral hip fractures (BHF, 31 cases), sacral fracture (SF, 5 cases), and sacral and hip fractures (SHF, 55 cases). The number of pelvic bone fragments per case ranged from 3 to 12 (5.7 ± 1.4).

The average voxel spacing is (0.83, 0.83, 0.89) mm. In cases where extra anatomies like abdomen or femoral shaft were included, the volumes were cropped to contain the pelvic region. The image dimensions average at (488, 426, 323). All the scans were de-indentified and converted to MHA format. 
The 150 cases are split into 100 for training, 20 for validation, and 30 for testing. To ensure a balanced representation in the test set, we employed stratified random sampling based on the five primary fracture types, so that the training and test sets have the same fracture type distribution.

\subsubsection{Annotation}
\label{sec:annotation}
The dataset is processed independently by two experienced annotators each with more than five years of medical imaging experience and a senior expert with more than 15 years of orthopedic surgical planning experience. 
The annotators were instructed to annotate three anatomical labels—left hipbone, right hipbone, and sacrum—as well as the bone fragments within each bone using the 3D Slicer software. 
The data annotation was structured into a four-step workflow:
\begin{enumerate}
    \item \textbf{Initial automatic segmentation:} We employ a pre-trained segmentation network based on the nnUNet framework to produce preliminary segmentations for the hipbones and sacrum. This network was trained on the CTPelvic1K dataset, with the majority of them not presenting any fractures~\cite{liu2021deep}.
    \item \textbf{Manual refinement of anatomical labels:} The initial anatomical segmentations were manually refined by trained annotators. This step involved removing any non-pelvic structures and restoring under-segmented regions, with particular attention paid to boundary regions near fracture surfaces where automatic segmentation is most prone to error. Refinement was performed using localized intensity thresholding and scissor-based region editing tools to ensure anatomically correct and topologically consistent bone surfaces.
    \item \textbf{Identification of fractured fragments:} Leveraging the refined anatomical labels, the annotators delineate individual fracture fragments using scissor-based region separation and slice-by-slice manual painting tools to ensure accurate fragment boundaries.
    \item \textbf{Expert adjudication:} The two independently produced annotations are then reviewed side-by-side by a senior expert, who selects the annotation that better reflects clinically realistic fracture separability. In cases where the two annotations are visually indistinguishable, selections are alternated across cases to avoid systematic bias. This adjudication step ensures label consistency while preserving clinical relevance.

\end{enumerate}

Before annotation began, the annotators underwent a training process, which included standardized written instructions and reference examples covering common fracture patterns. A key aspect of the refinement procedure involves determining whether an incomplete fracture (a partially fractured but non-displaced cortical discontinuity) should be labeled as one fragment or two fragments. We provide the following rule-of-thumb to guide this decision: If the cortical separation is expected to cause independent movement of the two regions during surgical reduction, the bone is labeled as separate fragments; If the two regions are expected to move together as a single rigid unit, the region is labeled as one fragment.
In addition, if a bone does not have any fracture, the only fragment corresponds to its anatomical mask. Fragments with a volume smaller than 500 mm$^3$ were omitted. The ground-truth labels were also saved in MHA format.

\begin{figure}[t]
\includegraphics[width=\columnwidth]{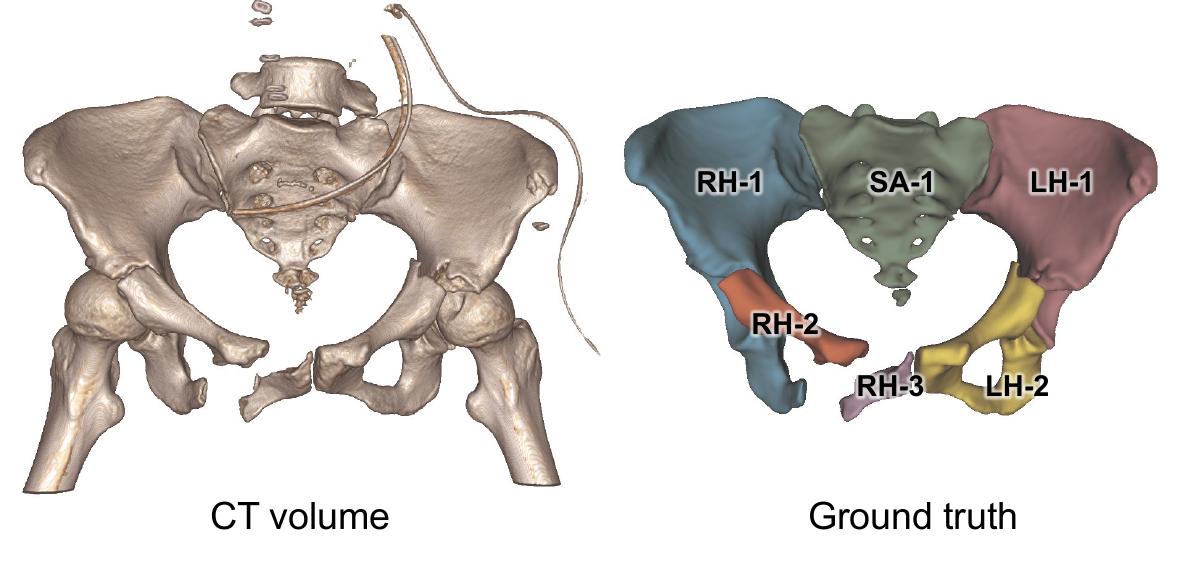}
\centering
\caption{Example pelvic CT scan and the annotated segmentation rendered in 3D view. The annotation follows a two-level structure: the first level indicates the anatomical origin, and the second level distinguishes individual fragment instances.}\label{fig:example_ct}
\end{figure}

Fig.~\ref{fig:example_ct} illustrates an example CT scan along with its annotated segmentation mask.  
In this instance segmentation task, the number of fragment instances varies across cases. The label indexing follows a two-level hierarchical structure: first, fragments are classified based on their anatomical origin, and second, they are ordered by fragment size. 
The CT training set is publicly available under a CC BY-NC-SA license\footnote{PENGWIN CT dataset: \href{https://doi.org/10.5281/zenodo.10927452}{{https://doi.org/10.5281/zenodo.10927452}}.}.  

\subsubsection{Generation of the X-Ray Dataset} 

\begin{figure}[t]
\includegraphics[width=\columnwidth]{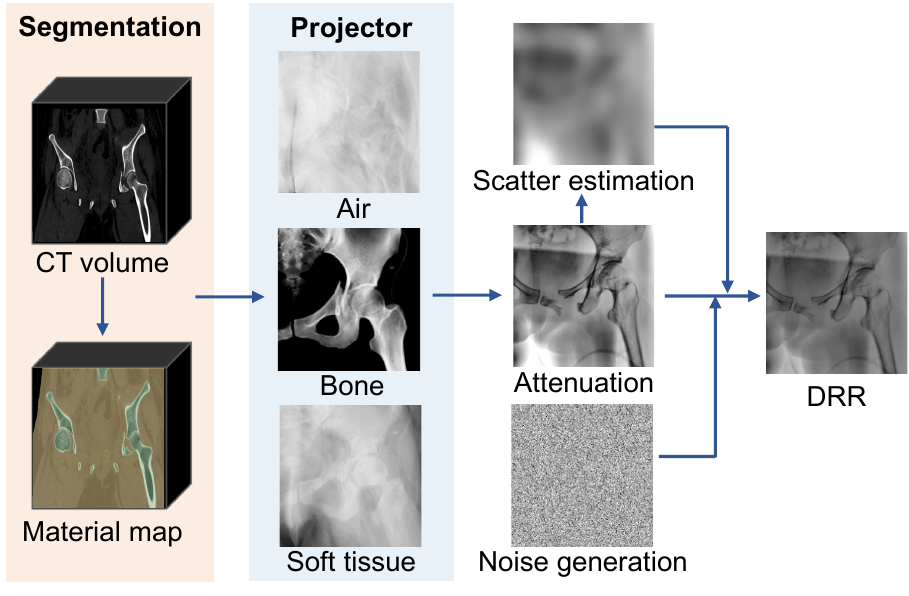}
\centering
\caption{The X-ray simulation process of DeepDRR.} \label{fig:DRR_process}
\end{figure}

\begin{figure}[t]
\includegraphics[width=\columnwidth]{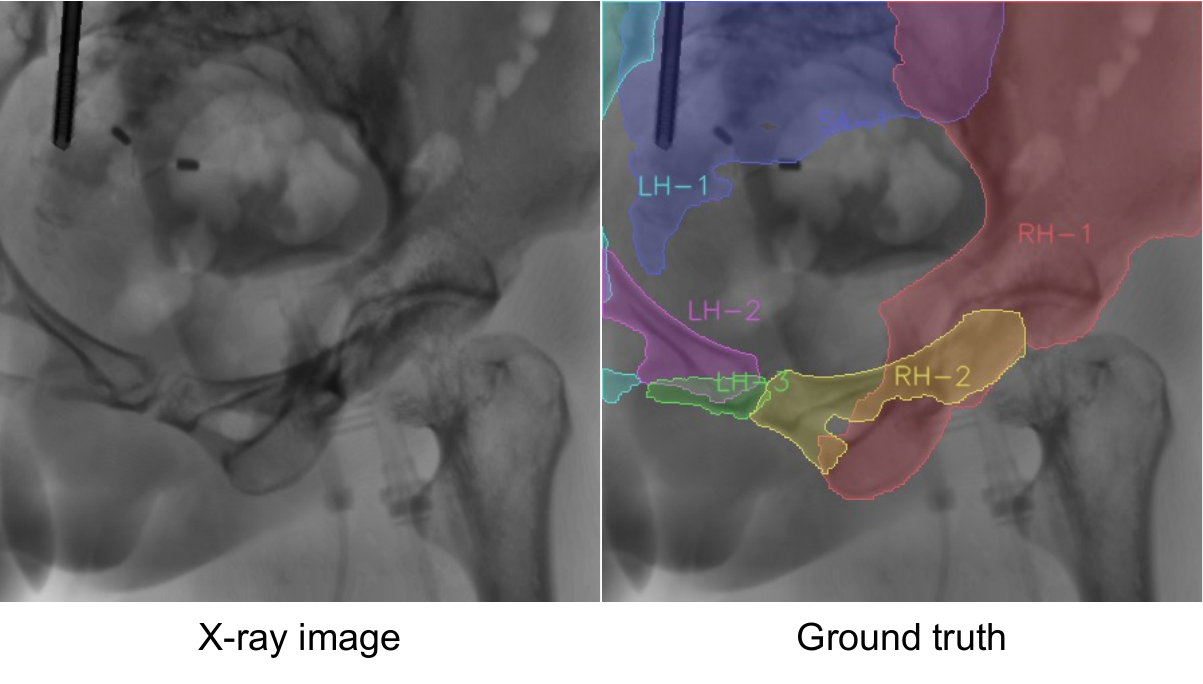}
\centering
\caption{Example simulated X-ray image and the annotation.} \label{fig:example_xray}
\end{figure}

The PENGWIN X-ray dataset comprises digitally reconstructed radiographs (DRRs) that simulate intraoperative C-arm X-ray images, including pelvic anatomy, fracture fragments, and surgical tools. To ensure robustness against surgical hardware interference, instrument poses were randomly varied across images.
The X-ray images were synthesized from CT data using the DeepDRR framework, which simulates realistic projections through a combination of analytic ray-tracing and learning-based scatter estimation and noise injection techniques~\cite{unberath2018deepdrr}. As shown in Fig.~\ref{fig:DRR_process}, the generation pipeline involves the following steps:

First, material decomposition is applied. A 3D convolutional network classifies every voxel in the CT volume as air, soft tissue or bone. The resulting material map preserves each tissue’s true spatial distribution and provides the physical input needed for X-ray simulation.

Second, forward projection is conducted using material- and spectrum-aware ray-tracing. Using this map, DeepDRR traces virtual X-ray beams from the source to the detector. For each detector pixel $u$, the transmitted intensity is
\begin{equation}
p(u)=\int_E p_0(E)\,
        \exp\!\Bigl(-\sum_{m\in M}
        (\mu/\rho)_m(E)\,
        \int_{\ell_u} \rho_m(x)\,d\ell\Bigr)\,dE,
\end{equation}
where $p_0(E)$ is the source spectrum, $(\mu/\rho)_m(E)$ the energy-dependent attenuation of material $m$, $\rho_m(x)$ its density along the ray $\ell_u$, and $M=\{\text{air},\ \text{soft tissue},\ \text{bone}\}$ is simulated materials. This step yields the primary, scatter-free projection.

Third, scatter estimation is performed. Real fluoroscopy contains low-frequency scatter that softens contrast. DeepDRR estimates this component with a lightweight ten-layer convolutional network trained on Monte-Carlo data. The predicted scatter image is added pixel-wise to the primary projection.

Finally, detector realism is achieved by adding two noise sources. Photon (quantum) noise is drawn from a Poisson distribution whose mean photon count,
\begin{equation}
N(u)=\sum_E \frac{p(E,u)}{E}\,N_0,
\end{equation}
is computed from the attenuated spectrum; $N_0$ is the incident photon fluence. Electronic read-out noise is added as correlated Gaussian noise. The result is a radiograph whose grain and contrast match clinical images.

The realism of DeepDRR has been validated in previous work~\cite{unberath2018deepdrr, gao2023synthetic}, which demonstrated that models trained solely on DeepDRRs generalized effectively to real fluoroscopy images without any domain adaptation.

The virtual C-arm parameters were randomized across the training set. Each view was sampled from a uniform distribution over a solid angle extending up to 60 degrees from the vertical axis, assuming a supine patient position. The image resolution was set to (448, 448) pixels. Ground-truth 3D segmentation masks were projected using the same camera geometry to generate corresponding 2D multi-label segmentation masks. Due to the projection nature of X-rays, overlapping bone fragments result in multi-label regions, making this a multi-label instance segmentation task. An example simulated X-ray and its ground-truth label are shown in Fig.~\ref{fig:example_xray}.

The simulated X-ray dataset was generated from the same 150 CT scans described above, following the same patient-wise train/validation/test split. For each CT scan in the training and validation sets, 400 X-ray images were simulated to provide sufficient diversity in view direction. For the test set, a reduced sampling of 20 X-ray images per CT scan was used, resulting in a total of 600 test images. This lower sampling density was chosen to avoid excessive redundancy among similar projection angles and to comply with computational constraints of the Grand Challenge platform, which executes containerized inference on a per-case basis.
The images were provided in binary Tag Image File Format
(TIFF), with multi-label segmentation masks encoded as single-channel binary TIFF images to comply with the challenge platform requirements. Additionally, the encoding and decoding scripts for handling segmentation files were made available to participants to ensure consistency in data processing. 
The X-ray training set is publicly available in an open repository under a CC BY-NC-SA license\footnote{PENGWIN X-ray dataset: \href{https://doi.org/10.5281/zenodo.10913195}{{https://doi.org/10.5281/zenodo.10913195}}.}. 

\subsection{Assessment Method}
\subsubsection{Evaluation Metrics}
The performance of submitted algorithms was evaluated using a set of quantitative metrics designed to assess both anatomical segmentation accuracy and fracture fragment delineation. The evaluation incorporated both overlap-based and distance-based measures to provide a comprehensive assessment.

For both the CT and X-ray tasks, segmentation performance was evaluated using intersection over union (IoU), the 95th percentile Hausdorff distance (HD95), and average symmetric surface distance (ASSD). IoU was calculated on solid masks and reflects the volumetric overlap between the predicted and ground-truth regions, serving as a general indicator of segmentation completeness. HD95 and ASSD were computed on surface contours to capture spatial discrepancies along fragment boundaries. While ASSD represents the average surface deviation and reflects typical alignment accuracy, HD95 quantifies the worst-case boundary error, excluding outliers. This is particularly important in clinical scenarios where large local inaccuracies near fracture lines or joint boundaries may compromise diagnosis or treatment planning. The adoption of these metrics aligns with recommendations from the Metrics Reloaded initiative and has been validated in prior benchmarking efforts \cite{faghani2022mitigating}.

Beyond algorithmic benchmarking, these metrics carry direct clinical relevance. Accurate estimation of fragment volume and displacement—reflected by IoU—is crucial for assessing fracture severity and planning fixation strategies. Surface-level metrics such as ASSD and HD95 are highly pertinent to the precision of virtual fracture reductions and implant alignment. Prior work has shown that segmentation errors propagate into increased inaccuracies in reduction pose estimation, underscoring the critical role of robust segmentation as a foundation for downstream surgical planning~\cite{liu2025preoperative}. 

IoU measures the overlap between the predicted segmentation mask $S_p$ and the ground truth mask $S_g$:
\begin{equation}
\text{IoU} = \frac{|S_p \cap S_g|}{|S_p \cup S_g|}.
\end{equation}

HD95 quantifies the worst-case segmentation error by measuring the 95th percentile of the distances between the predicted and ground-truth segmentation contours:
\begin{equation}
\text{HD}_{95} = \max\left\{ \sup_{x \in C_p} \inf_{y \in C_g} d(x, y), \sup_{y \in C_g} \inf_{x \in C_p} d(x, y) \right\}_{95\%},
\end{equation}
where $C_p$ and $C_g$ denote the surfaces (contours) of the predicted and ground-truth segmentations, respectively, and $d(x, y)$ is the Euclidean distance between points $x$ and $y$.

ASSD computes the average bidirectional distance between the predicted and ground truth segmentation contours, ensuring a smooth and accurate segmentation. It is defined as:
\begin{equation}
\text{ASSD} = \frac{1}{|C_p| + |C_g|} ( \sum_{x \in C_p} \inf_{y \in C_g} d(x, y) + \sum_{y \in C_g} \inf_{x \in C_p} d(x, y) ),
\end{equation}
where distances are computed in both directions.

Evaluation was performed on both fracture segmentation and anatomical segmentation results. Fracture segmentation performance was assessed by comparing each matched fragment instance with its corresponding ground truth. Anatomical segmentation performance was evaluated by merging the predicted labels for all bone structures and comparing them to the ground-truth anatomical labels of the left hipbone, right hipbone, and sacrum.
In the case of fragment instance segmentation, each ground-truth fragment was matched to the predicted fragment with the highest IoU, ensuring a one-to-one correspondence in performance assessment. In cases where a fragment was entirely missing in the prediction, HD95 and ASSD were set to the diameter and radius, respectively, of the ground-truth fragment’s bounding sphere.
As a result, six evaluation metrics were used: IoU-F, HD95-F, and ASSD-F for fracture segmentation and IoU-A, HD95-A, and ASSD-A for anatomical segmentation. This metric scheme was applied to both tasks and has been made publicly available\footnote{Evaluation code: \href{https://github.com/YzzLiu/PENGWIN-example-algorithm}{{https://github.com/YzzLiu/PENGWIN-example-algorithm}}.}.

In addition to accuracy metrics, we also report the container runtime as an indicator of efficiency. Note that this runtime includes not only the model inference time but also the overhead from container initialization, image loading, and result saving.

\subsubsection{Ranking Scheme}
The ranking procedure consisted of the following steps:

\begin{enumerate}
    \item \textbf{Metric calculation:} For each individual test case, the accuracy metrics were computed for both fracture segmentation and anatomical segmentation. 
    \item \textbf{Independent ranking:} For each submission, the average performance across all test cases was calculated for each metric. Submissions were then ranked independently based on their performance on each metric.
    \item \textbf{Final ranking:} The final rank of each submission was determined by computing the average of its ranks across all evaluation metrics. This approach ensured that submissions were assessed holistically rather than being overly influenced by a single metric.
    \item \textbf{Tie-breaking:} In the event of a tie, the container execution time of the submission was used as a tie-breaker, with faster models receiving a higher rank.
\end{enumerate}

\section{Results}
\subsection{Participation}
The challenge attracted interest from the medical imaging and machine learning communities. A total of 239 participants from 25 countries registered for the challenge, gaining access to the dataset. Among them, 26 registered teams were formed.

During the preliminary development phase, participants actively engaged with the challenge, resulting in 256 submissions for the CT task and 84 submissions for the X-ray task.

During the final test phase, 11 teams from five countries submitted models for the CT task, while five teams from three countries participated in the X-ray task. Among them, two teams chose to complete both tasks.

\subsection{Algorithm Summary}
In this section, we summarize the methods employed by the top five performing teams in the CT task and the top four performing teams in the X-ray task, as subsequent teams exhibited significantly larger performance gaps in comparison.

\subsubsection{CT Segmentation}
\begin{figure*}[t]
\centering
\includegraphics[width=0.9\textwidth]{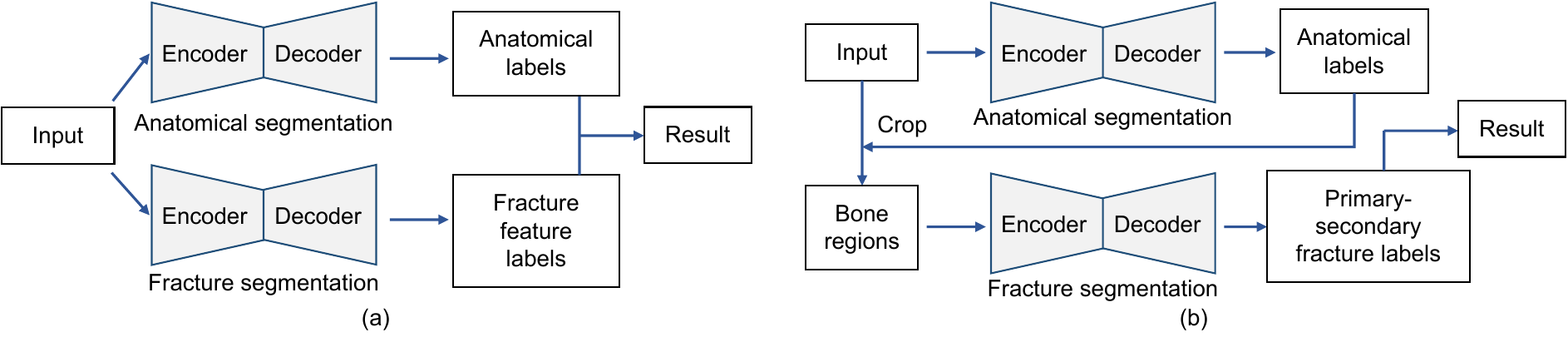}
\caption{Different fracture segmentation pipelines in the CT task. (a) Boundary-core decomposition scheme adopted by MIC-DKFZ, MedApp-AGH, and Sano. (b) Primary-secondary labeling scheme adopted by SMILE and MedIG.} \label{fig:Fig_scheme_ct}
\end{figure*}

\begin{table*}[t]
    \caption{Summary of the top-performing participating algorithms for CT segmentation. Spatial augmentation includes random scaling, mirroring, and flipping. Intensity augmentation includes noising, blurring, and histogram adjustments.}\label{tab:algorithm_CT}
    \setlength{\arrayrulewidth}{0.1mm}
    \centering
    \resizebox{\textwidth}{!}{%
    \begin{tabular}{llllllll}
        \hline
        Team & External resource & Anatomical network & Fracture network & Pre-processing & Augmentation & Interpolation & Loss \\  
        \hline
        MIC-DKFZ & - & UNet with residual & UNet with residual & standard nnUNet & spatial, intensity, low-res & cubic; anisotropic & Dice, CE \\  
        SMILE & - & UNet & UNet & standard nnUNet & spatial, cropping & linear; isotropic & Dice, CE \\  
        MedIG & TotalSegmentator & Pre-trained UNet & NexToU & standard nnUNet & spatial, intensity, low-res & cubic; anisotropic & Dice, CE  \\  
        MedApp-AGH & FracSegNet & UNet & UNet & standard nnUNet & spatial & cubic; isotropic & Dice, CE  \\  
        Sano & - & UNet & UNet & normalize, reorient & spatial, intensity & linear; isotropic & Dice, CE \\  
        \hline
    \end{tabular}
    } 
\end{table*}

\paragraph{\textbf{MIC-DKFZ} (1st place, Johannsen et al.)} 

The authors formulated the task as two semantic segmentation problems. The first network performed anatomical segmentation of the three bone regions. The second network predicted the adaptive border boundary core (ABBC) representation, which consists of four semantic classes: background, boundary, border, and core. The core and boundary classes together define individual fragments, with the boundary class acting as a separator between the core and the background or adjacent fragments. The authors introduced a novel dynamic boundary thickness adjustment using the medial axis transform, ensuring that thin fragments remain connected within the core region. 
Given a prediction, connected component analysis was applied to the core class to extract initial fragment instances. The boundary class was then merged with the nearest seed. To ensure anatomical consistency, fragment instances were adjusted by splitting them along anatomical borders and merging smaller fragments with the closest anatomically correct instance. Additionally, the border class was used to refine the segmentation by modifying fragments through split-merge operations, ensuring alignment with the network predictions.\footnote{MIC-DKFZ CT algorithm: \href{https://codebase.helmholtz.cloud/hi-dkfz/applied-computer-vision-lab/challenges/abbc}{{https://codebase.helmholtz.cloud/hi-dkfz/applied-computer-vision-lab/challenges/abbc}}.}

\paragraph{\textbf{SMILE} (2nd place, Yue et al.)} 
The authors proposed a two-stage segmentation framework.
The first stage focused on localizing the three bone regions. To achieve this, the authors trained a U-Net model on downsampled CT scans, which reduced the computational costs and provided a global context that prevented excessive false positives when processing smaller patches. The localized regions were then refined at full resolution for accurate anatomical segmentation. 
The second stage focused on fracture segmentation within each of the three bone regions. The authors categorized bone fragments into primary and secondary components. After obtaining initial predictions, connected component analysis was applied to reassign secondary fragment labels in accordance with evaluation rules.
The final segmentation result was generated by combining the outputs from both stages. A final refinement step ensured that the fracture segmentation was aligned with the foreground mask from the first stage.\footnote{SMILE CT algorithm: \href{https://github.com/yuepeiyan/PENGWIN\_Challenge}{{https://github.com/yuepeiyan/PENGWIN_Challenge}}.}

\paragraph{\textbf{MedIG} (3rd place, Pan et al.)}
The authors employed a two-stage segmentation pipeline to accurately localize and segment fractured bone fragments.
In the first stage, the TotalSegmentator~\cite{wasserthal2023totalsegmentator} model was used to localize the pelvic region, after which the three bone regions were cropped from the original CT scans.
In the second stage, the NexToU model was applied to segment primary and secondary fragments within each cropped bone region. The largest connected component was extracted as the primary fragment, and the remaining secondary fragments were expected to be naturally separated. A connected component analysis step was used to isolate individual secondary fragments.\footnote{MedIG CT algorithm: \href{https://github.com/pzhhhhh2263/MICCAI-Challenge-PENGWIN2024}{{https://github.com/pzhhhhh2263/MICCAI-Challenge-PENGWIN2024}}.}

\paragraph{\textbf{MedApp-AGH} (4th place, Jurgas et al.)} 
The authors developed a refinement-based segmentation pipeline, integrating anatomical segmentation with fracture surface segmentation to ensure anatomically consistent fracture identification.
The approach utilized two deep networks. The first network, adapted from the FracSegNet study~\cite{liu2023pelvic}, was used for segmenting the three bone regions. The second network, trained on fracture surfaces extracted from the PENGWIN dataset, identified fracture boundaries.
For training, the authors generated a complete anatomical segmentation of the pelvis by merging all bone fragments. They then extracted contours of both individual fragments and the anatomical segmentation, removing overlapping edges to isolate fracture surfaces—where fractured bones were in contact. The fracture network was trained to predict these fracture surfaces from CT scans.
In post-processing, segmentations from both networks were merged. A continuity analysis was then performed to separate individual fractures, with fragments labeled according to their anatomical overlap.\footnote{MedApp-AGH CT algorithm: \href{https://github.com/Jarartur/pengwin-challenge-submission}{{https://github.com/Jarartur/pengwin-challenge-submission}}.}

\paragraph{\textbf{Sano} (5th place, Płotka et al.)} 
The authors proposed PeFreCT, which employed two UNet models. The first network performed anatomical segmentation, while the second network focused on fracture segmentation, detecting fracture regions based on textural features. For training the anatomical network, the authors cropped the CT scans by first detecting the body using radio-density thresholding and removing artifacts with morphological operations. The dataset was doubled by flipping images along the sagittal plane and mirroring hipbone labels. For training the fracture network, they precomputed the fracture region mask by dilating the masks of all fragments and identifying overlapping borders to form a final binary fracture label. Since fracture detection based on texture is predominantly local, a (64, 64, 64) patching scheme was applied for both training and inference to enhance efficiency. In post-processing, the binary fracture mask were subtracted from anatomical masks, allowing for fragment separation with connected component analysis. Finally, the fragment masks were expanded to include the fracture regions, using the watershed algorithm guided by raw fracture probability maps to precisely delineate the borders between colliding fragments.

\paragraph{\textbf{Summary}} 
All five teams employed a two-stage approach, first performing anatomical segmentation before segmenting fragments. As shown in Fig.~\ref{fig:Fig_scheme_ct}, the key variations among their methods lay in the fracture segmentation step—whether to extract individual fragments directly or to first detect border regions and infer separation.  
In Table \ref{tab:algorithm_CT}, we summarize some of the
main components of the algorithms, including network architecture, pre-processing, and training schemes. 
A detailed discussion of methodology is included in Sec.~\ref{discussion_scheme_ct}.

\subsubsection{X-Ray Segmentation}
\begin{figure*}[t]
\centering
\includegraphics[width=0.9\textwidth]{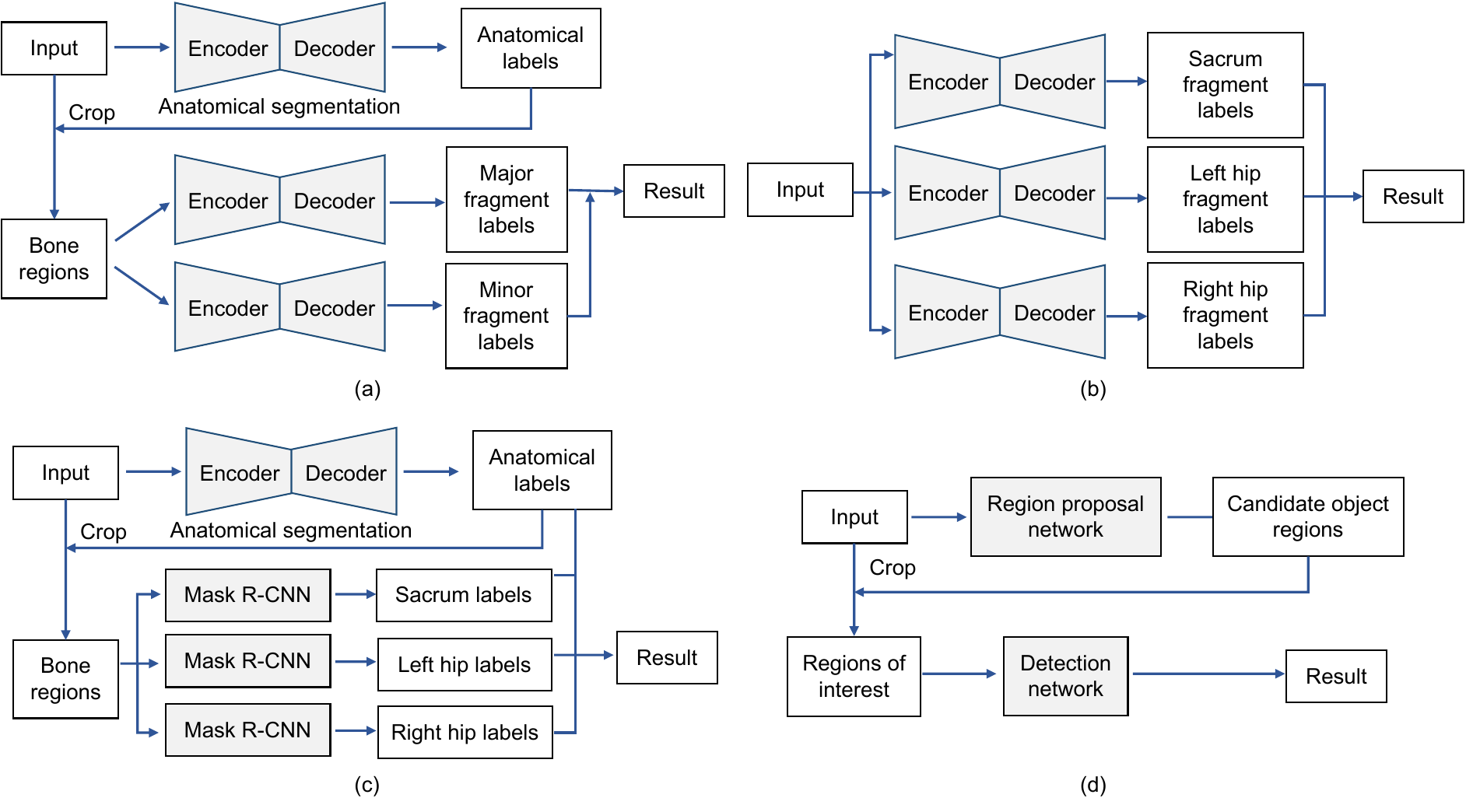}
\caption{Different fracture segmentation pipelines in the X-ray task. (a) Bone extraction followed by primary-secondary segmentation by SMILE. (b) Direct semantic segmentation for the three bones in parallel by MedIG. (c) Bone extraction followed by instance segmentation by LME. (d) Direct instance segmentation with anatomical classification by luckyjing.} \label{fig:Fig_scheme_xray}
\end{figure*}

\begin{table*}[t]
    \caption{Summary of the top-performing participating algorithms for X-ray segmentation. Spatial augmentation includes random scaling, mirroring, and flipping. Intensity augmentation includes noising, blurring, and histogram adjustments.}\label{tab:algorithm_xray}
    \setlength{\arrayrulewidth}{0.1mm}
    \centering
    \resizebox{\textwidth}{!}{%
    \begin{tabular}{lllllll}
        \hline
        Team & Overall scheme & External resource & Network & Pre-processing & Augmentation & Loss \\  
        \hline
        SMILE & Bone extraction followed by primary-secondary segmentation & - & UNet & standard nnUNet & spatial, crop & Dice, CE \\  
        MedIG & Direct semantic segmentation for the three bones in parallel & ImageNet & Unet++ with RegNetY-160 & crop, normalize & - & Dice, CE, Focal Tversky \\  
        LME & Bone extraction followed by instance segmentation & COCO & Swin UNETR; Mask R-CNN & pad & intensity & Dice \\  
        luckyjing & Direct instance segmentation with anatomical classification & - & Mask R-CNN with ResNet50 & normalize & flip, crop, pad & CE, smooth L1, IoU \\  
        \hline
    \end{tabular}    } 
\end{table*}

\paragraph{\textbf{SMILE} (1st place, Yue et al.)} 
The authors addressed the limitations of semantic segmentation in handling overlapping anatomical structures in X-rays by developing a two-stage hierarchical strategy similar to their CT solution.
In the first stage, three dedicated U-Net models were trained to segment the three bone regions separately, allowing for focused substructure analysis within each area. 
In the second stage, a hierarchical fragment classification approach was implemented, adapted from the authors’ CT segmentation method, with modifications specific to X-ray imaging. Separate models were employed to distinguish primary and secondary fracture fragments, mitigating the overlapping structures in X-rays. While this approach effectively reduced primary-secondary overlap, challenges remained in cases where multiple secondary fragments overlapped.
For post-processing, connected component analysis was applied to relabel secondary fragments, while primary fragment predictions were refined through consistency checks with the localization masks from the first stage, ensuring that only validated foreground pixels were retained.\footnote{SMILE X-ray algorithm: \href{https://github.com/yuepeiyan/PENGWIN\_Challenge}{{https://github.com/yuepeiyan/PENGWIN_Challenge}}.}

\paragraph{\textbf{MedIG} (2nd place, Pan et al.)} 
The authors employed a multi-model approach to process each of the three bone regions in X-rays. Each model generated a multi-channel 3D tensor as output. The first channel represents the primary fragment, while the remaining nine channels corresponds to secondary fragments. The segmentation results from the three models were then combined to construct the final output. 
To optimize model performance, the authors applied as loss function an unweighted combination of Dice loss, binary CE loss, and focal Tversky loss. A pre-trained ImageNet model was used to initialize the network, enhancing feature extraction capabilities. The network architecture was based on a UNet++ backbone with a RegNetY-160 encoder, which provided a balance between spatial resolution and computational efficiency.\footnote{MedIG X-ray algorithm: \href{https://github.com/pzhhhhh2263/MICCAI-Challenge-PENGWIN2024}{{https://github.com/pzhhhhh2263/MICCAI-Challenge-PENGWIN2024}}.}

\paragraph{\textbf{LME} (3rd place, Liu et al.)} 
The authors proposed a category and fragment segmentation (CFS) framework. First, three separate networks, each based on the Swin-UNETR architecture, were used to segment the three bone regions. Then, a fragment segmentation network was applied to isolate the bone fragments within each region.
The fragment segmentation step employed three separate Mask R-CNN-based networks, which performed both segmentation and object detection through bounding box prediction and classification. These networks took both the image and the binary anatomical mask as input, allowing the model to focus on specific bone regions while reducing interference from surrounding structures. During inference, the network predicted multiple bounding boxes, each associated with a confidence score. Only bounding boxes with a confidence score above 0.8 were retained for further analysis. In post-processing, the fragment segmentation masks were multiplied by the anatomical masks from the first step to ensure spatial consistency.

\paragraph{\textbf{luckyjing} (4th place, Ma et al.)} 
The authors employed Mask R-CNN with a ResNet50 backbone for instance segmentation. The model consists of two main stages: a region proposal network that generates candidate object regions, and a detection network that refines these regions while performing anatomical classification, bounding-box regression, and mask prediction. Mask R-CNN was employed to predict pixel-wise binary masks for each region of interest (RoI). To maintain spatial alignment between RoIs and extracted features, the network replaced RoI Pooling with RoI Align, which eliminates quantization errors by using bilinear interpolation instead of discrete binning. This significantly improved mask prediction accuracy, particularly for small objects. To address the specific challenges of pelvic fragment segmentation, the loss weights between classification and segmentation were carefully adjusted during training. Additionally, a structured loss function combining IoU and cross-entropy loss was incorporated, enhancing the model’s robustness and segmentation quality in occluded scenarios.

\paragraph{\textbf{Summary}} 
As shown in Fig.~\ref{fig:Fig_scheme_xray}, the solutions in the X-ray task exhibited greater diversity in their overall approach than the CT task. The main components of the algorithms are summarized in Table \ref{tab:algorithm_xray} and further discussed in Sec.~\ref{discussion_scheme_xray}.

\subsection{Performance Analysis}
\subsubsection{CT Segmentation Performance}

\begin{figure*}[t]
\centering
\includegraphics[width=\textwidth]{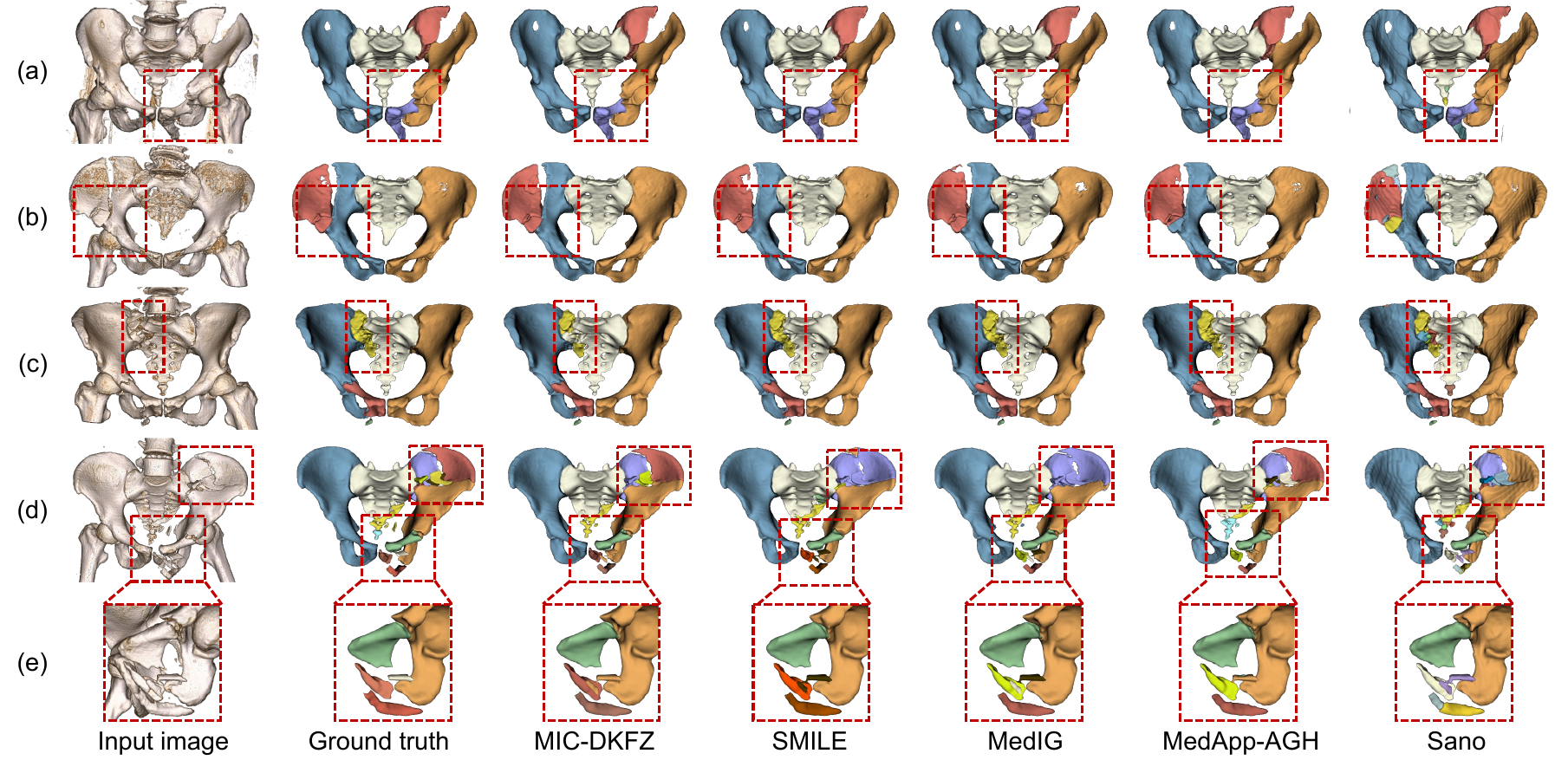}
\caption{Example CT segmentation results from the top-performing teams. Four representative cases are shown in (a)-(d), illustrating varying levels of complexity from easy to difficult. A complex fractured ramus pubicus region is shown in (e).
} \label{fig:Fig_Qualitative3D}
\end{figure*}

Example CT segmentation results are shown in Fig.~\ref{fig:Fig_Qualitative3D}. The performance metrics of each algorithm submitted to the final test are summarized in Table~\ref{tab:result_CT}. 
In addition, we evaluate the performance of our previously proposed CT semantic segmentation method, FracSegNet, described in Sec.~\ref{sec:challenges}, as a retrospective baseline. 

\begin{table*}[t]
    \caption{Quantitative results for the final test phase of the CT task. HD95 and ASSD values are reported in millimeters. Container runtime is reported in seconds. The best scores for each metric are shown in bold.}\label{tab:result_CT}
    \centering
    \begin{tabular}{lcccccccc}
        \hline
        Team & Rank ↓ & IoU-F ↑ & HD95-F ↓ & ASSD-F ↓ & IoU-A ↑ & HD95-A ↓ & ASSD-A ↓  & Time ↓ \\ 
        \hline
        MIC-DKFZ & 1 & \textbf{0.9296} & 5.866 & 1.843 & \textbf{0.9810} & \textbf{2.368} & \textbf{1.290} & 510.4 \\  
        SMILE & 2 & 0.9096 & \textbf{5.562} & \textbf{1.663} & 0.9802 & 2.410 & 1.321 & 333.1 \\  
        MedIG & 3 & 0.9084 & 6.207 & 1.680 & 0.9798 & 2.452 & 1.298 & 257.4 \\  
        MedApp-AGH & 4 & 0.9049 & 6.899 & 2.007 & 0.9783 & 2.398 & 1.385 & 617.2 \\  
        Sano & 5 & 0.8802 & 6.866 & 1.697 & 0.9491 & 2.522 & 1.343 & 308.1 \\  
        Lee-SKKU-LG & 6 & 0.8079 & 17.449 & 3.612 & 0.9718 & 5.283 & 1.576 & 535.5 \\  
        CLone & 7 & 0.6339 & 46.486 & 8.899 & 0.9332 & 16.119 & 2.730 & \textbf{160.0} \\  
        Rzjs & 8 & 0.6787 & 48.742 & 7.963 & 0.9287 & 18.322 & 2.806 & 343.5 \\  
        Zeng-SJTU & 9 & 0.5423 & 46.457 & 17.757 & 0.6235 & 37.966 & 12.306 & 206.9 \\  
        Zou-SZU & 10 & 0.5944 & 86.892 & 18.090 & 0.6708 & 115.814 & 22.005 & 360.7 \\  
        watsons & 11 & 0.4940 & 83.746 & 19.186 & 0.6165 & 90.970 & 16.478 & 255.4 \\  
        \hline
        Baseline & - & 0.9003 & 6.967 & 2.178 & 0.9804 & 2.423 & 1.299 & - \\  
        Inter-observer & - & 0.9848 & 1.317 & 0.306 & 0.9996 & 1.069 & 0.474 & - \\      
        \hline
        \end{tabular}
\end{table*}

\subsubsection{X-Ray Segmentation Performance}

\begin{figure*}[t]
\centering
\includegraphics[width=\textwidth]{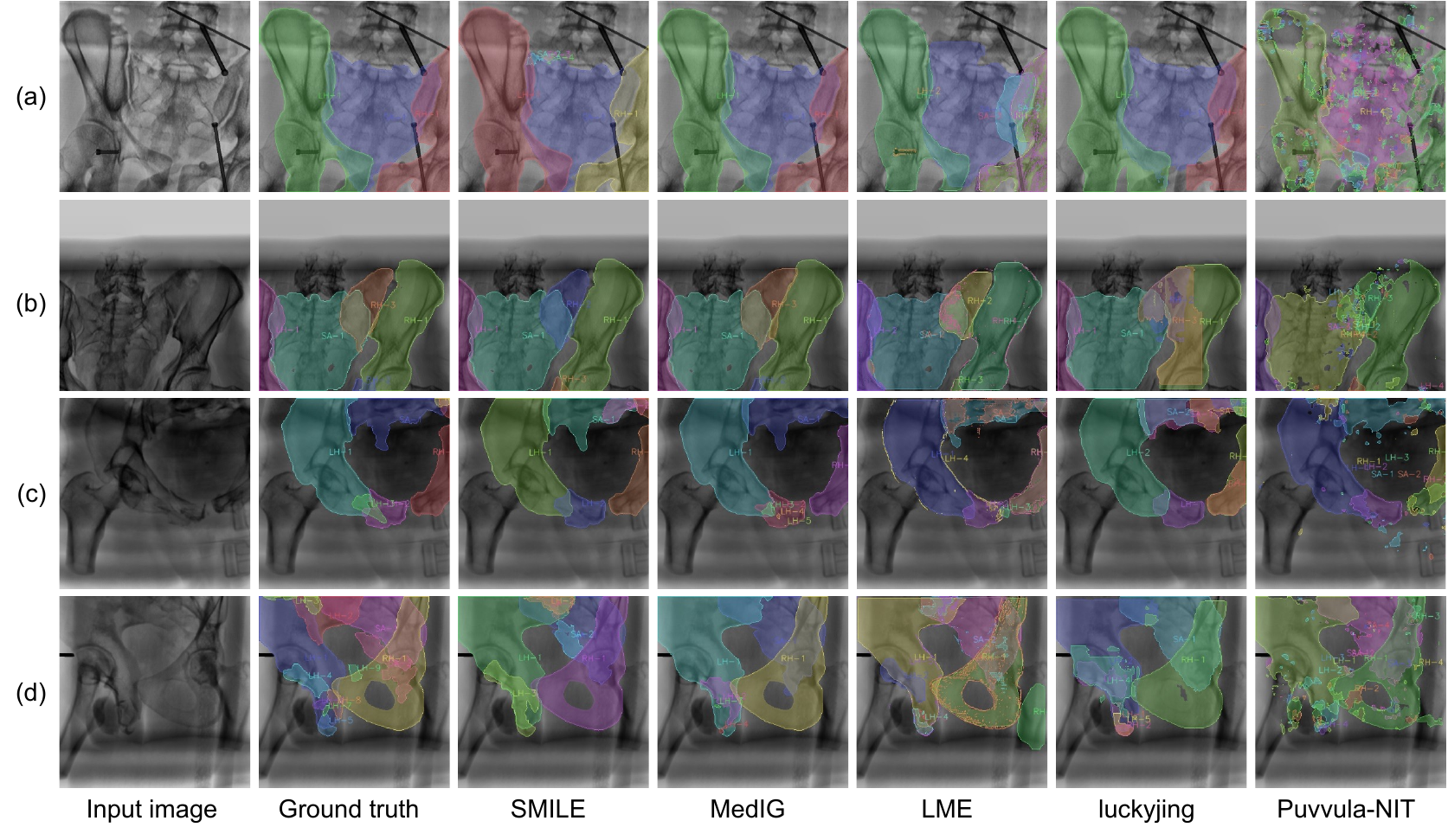}
\caption{Example X-ray segmentation results from the top-performing teams. Four representative cases are shown in (a)-(d), illustrating varying levels of complexity from easy to difficult.} \label{fig:Fig_Qualitative2D}
\end{figure*}

\begin{figure*}[t]
\centering
\includegraphics[width=0.9\textwidth]{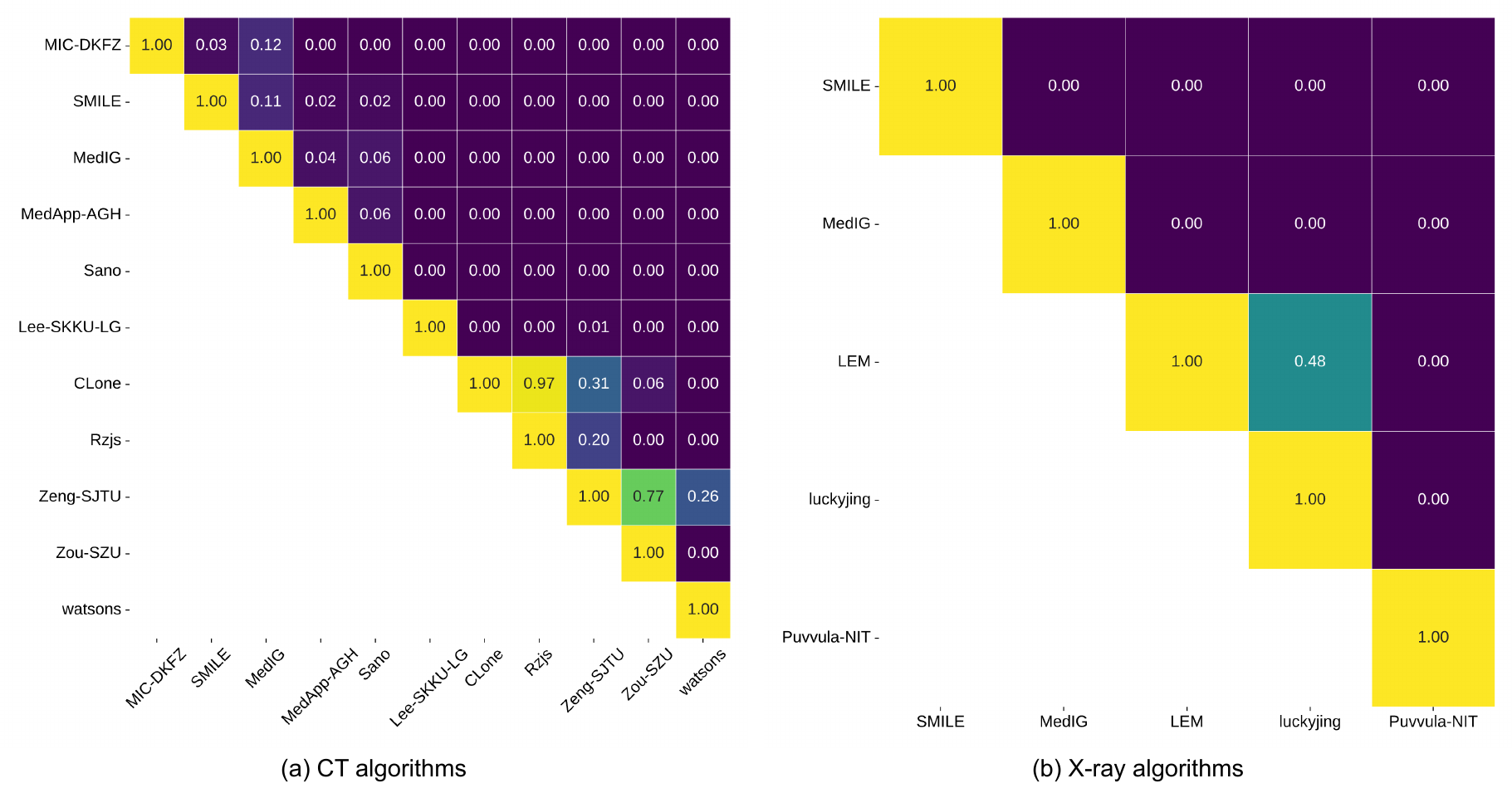}
\caption{Significance matrices showing the \textit{p}-values of the participating algorithms using one-sided Wilcoxon signed-rank tests.} \label{fig:significance}
\end{figure*}

Example X-ray segmentation results are shown in Fig.~\ref{fig:Fig_Qualitative2D}. The performance metrics of each algorithm submitted to the final test are summarized in Table~\ref{tab:result_Xray}.
To assess the statistical significance of pairwise differences between competing teams, we conducted a one-sided Wilcoxon signed-rank test to determine whether the IoU-F scores of one team were statistically significantly higher than those of another. 
The significance matrices are shown in Fig.~\ref{fig:significance}. A \textit{p}-value below 0.05 was considered statistically significant. 

\begin{table*}[t]
    \caption{Quantitative results for the final test phase of the X-ray task. HD95 and ASSD values are reported in pixel unit. Container runtime is reported in seconds. The best scores for each metric are shown in bold.}\label{tab:result_Xray}
    \centering
    \begin{tabular}{lcccccccc}
        \hline
        Team & Rank ↓ & IoU-F ↑ & HD95-F ↓ & ASSD-F ↓ & IoU-A ↑ & HD95-A ↓ & ASSD-A ↓ & Time ↓ \\ 
        \hline
        SMILE & 1 & \textbf{0.7736} & \textbf{37.366} & \textbf{8.545} & 0.9238 & \textbf{13.299} & \textbf{2.222} & 207.4 \\  
        MedIG & 2 & 0.7640 & 40.931 & 9.043 & \textbf{0.9268} & 15.979 & 2.462 & 168.5 \\  
        LME & 3 & 0.7150 & 42.986 & 10.355 & 0.8760 & 21.474 & 4.049 & 210.9 \\  
        luckyjing & 4 & 0.7145 & 45.752 & 10.398 & 0.8125 & 46.060 & 8.877 & 141.3 \\  
        Puvvula-NIT & 5 & 0.5506 & 158.171 & 28.092 & 0.7599 & 150.135 & 23.778 & \textbf{140.8}\\  
        \hline
    \end{tabular}
\end{table*}

\subsection{Inter-Observer Agreement}
\label{inter-observer}

In order to assess the quality of the ground truth, as well as to compare the algorithm's performance against human experts', we perform a retrospective analysis to assess inter-observer variability by manually annotating the CT test set with an additional independent annotator, using the same protocol described in Sec.~\ref{sec:annotation}. 

The assessment results are also summarized in Table~\ref{tab:result_CT}. The high agreement between annotators confirmed the reliability of the ground truth. The primary source of variation stemmed from differences in determining whether an incomplete fracture should be labeled as two separate fragments or a single fragment in certain cases.

For the X-ray task, however, a direct evaluation of inter-observer variability is not feasible, as the ground-truth labels is automatically generated through projection, and manual annotation of the projected labels proved exceedingly difficult.

\subsection{Impact of Fragment Size and Count on Segmentation}

\begin{figure*}[t]
\centering
\includegraphics[width=\textwidth]{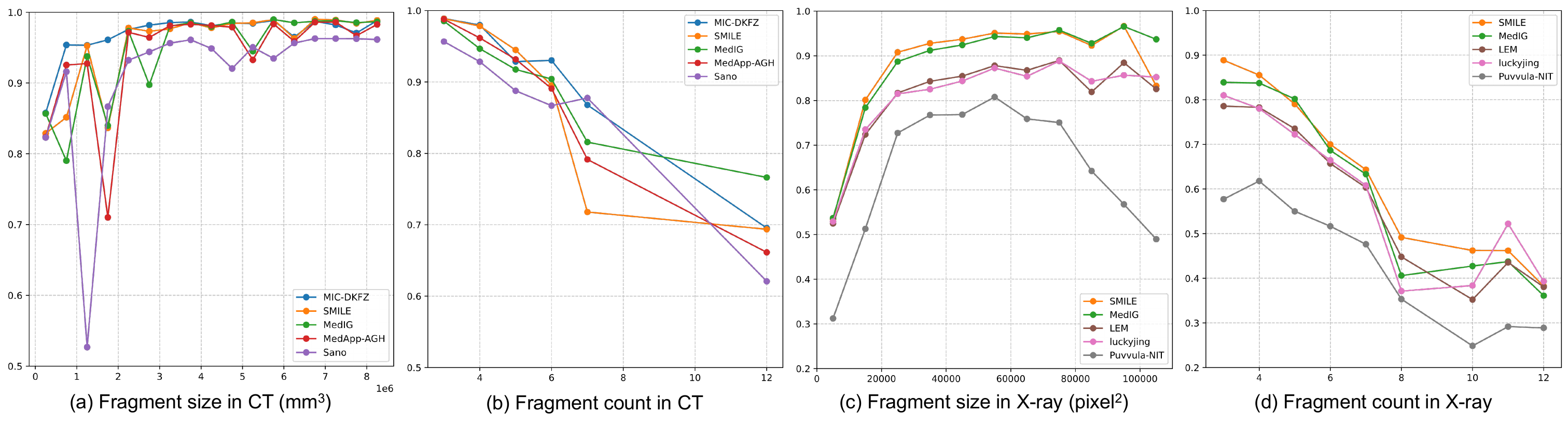}
\caption{Impact of fragment size and count on the IoU-F metric for the top five performing algorithms in both tasks. In (a) and (c), fragment sizes are binned in intervals of 500,000 mm\textsuperscript{3} for CT and 10,000 pixel\textsuperscript{2} for X-ray to enhance visual clarity.} \label{fig:fragmentVsIoU}
\end{figure*}

We investigated the impact of fragment size and fragment count on segmentation performance, using results from the top five performing algorithms in both the CT and X-ray tasks. The findings are illustrated in Fig.~\ref{fig:fragmentVsIoU}.

Overall, IoU-F scores decrease as fragment size decreases and fragment count increases, indicating the increased difficulty in accurately isolating or delineating small fragments, particularly in complex fracture scenarios.

\subsection{False-Positive Analysis}
\label{FP_analysis}

\begin{table}[t]
\centering
\caption{False-positive predictions in the final test phase. The mean and standard deviation of unmatched fragment counts are reported. The upper and lower sections show results for the CT and X-ray tasks, respectively.}
\label{tab:false_positive}
\begin{tabular}{lccc}
\hline
Team & Rank & Mean & Std \\
\hline
MIC-DKFZ & 1 & 0.53 & 0.78 \\
SMILE & 2 & 0.20 & 0.41 \\
MedIG & 3 & 0.93 & 0.94 \\
MedApp-AGH & 4 & 0.50 & 0.68 \\
Sano & 5 & 19.23 & 3.74 \\
\hline
SMILE & 1 & 0.50 & 0.95 \\
MedIG & 2 & 0.42 & 0.67 \\
LME & 3 & 3.12 & 1.13 \\
luckyjing & 4 & 2.11 & 1.64 \\
\hline
\end{tabular}
\end{table}

In our evaluation framework, false-positive (FP) fragments were not explicitly penalized at the instance level, but were instead indirectly reflected through the anatomical segmentation metrics. To better understand the prevalence and cause of FPs, we additionally computed the number of unmatched predicted fragments for each participating algorithm with respect to the ground truth, regardless of its size or shape. This analysis provides insight into over-segmentation tendencies and complements the primary evaluation metrics. The results are summarized in Table~\ref{tab:false_positive}. 

In the CT task, SMILE achieved the lowest FP rate, attributable to its coarse-to-fine anatomical segmentation strategy, which effectively excluded non-pelvic bony structures, and the use of a primary-secondary representation scheme that tended to under-segment small fragments rather than over-segment them. The algorithm developed by Sano exhibited a higher FP rate due to its tendency to over-segment small fragments. However, a portion of these fragments actually appeared in plausible locations—such as regions with thin, deformed cortical bone—and could be eliminated through post-processing with size-based filtering. Other sources of FP predictions included erroneous anatomical segmentation that mistakenly included patients’ hands or the CT table, as well as over-detection in regions where fractures commonly occur.

In the X-ray task, the primary-secondary semantic segmentation approach adopted by SMILE and MedIG resulted in fewer FPs compared to the instance segmentation strategies employed by LME and luckyjing. While instance-based methods hold potential for more precise identification of small fragments, they tend to be less robust in practice.

\subsection{Ranking Robustness}

\begin{figure*}[t]
\centering
\includegraphics[width=0.9\textwidth]{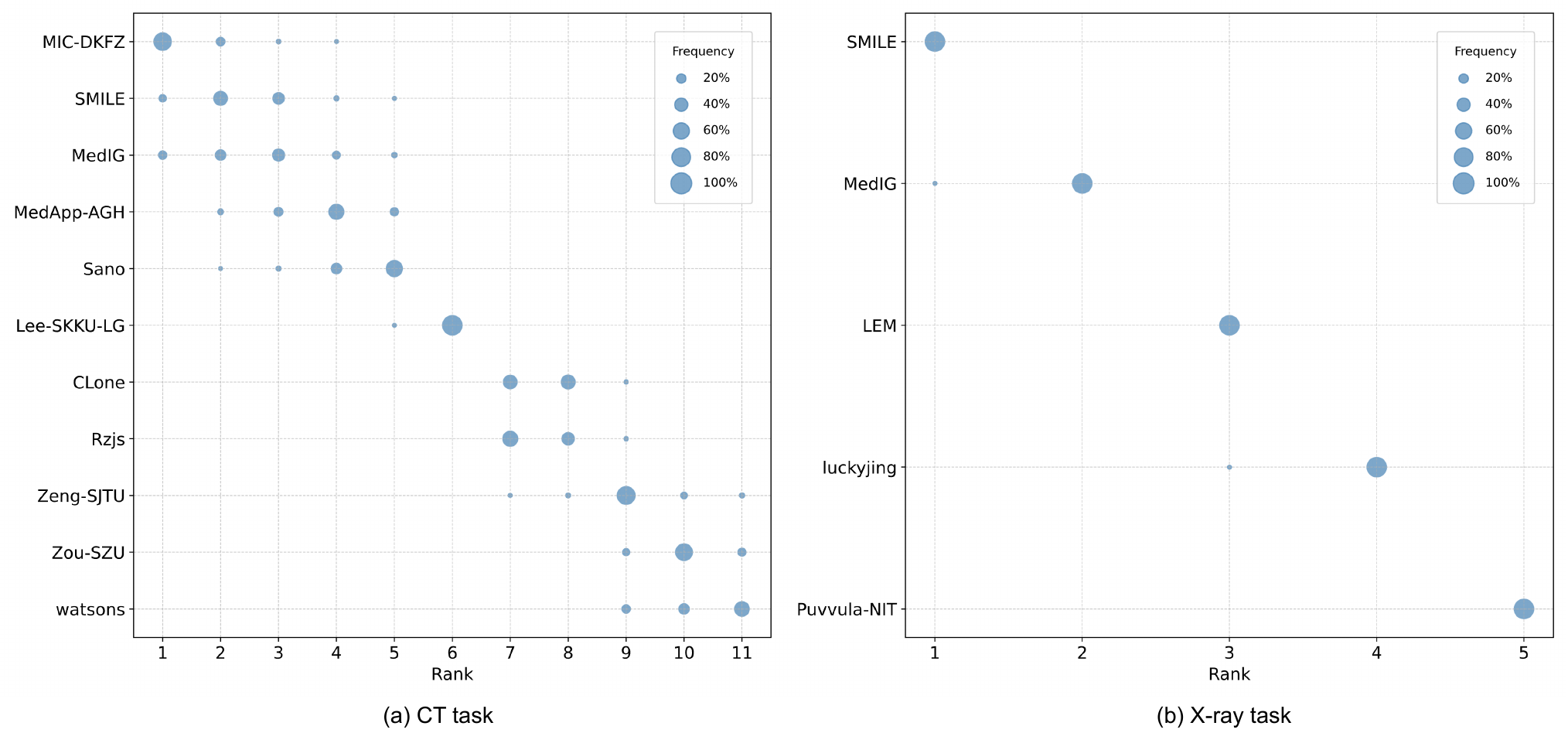}
\caption{Bootstrap-based stability analysis of team rankings in both tasks. Each dot represents the frequency with which a team achieved a specific rank across 1,000 bootstrap samples of the test set. Dot size is proportional to the frequency of that rank being assigned.} \label{fig:ranking}
\end{figure*}

To evaluate the robustness of the final team rankings against variations in the test set, we performed a bootstrap analysis using 1,000 randomly resampled replacements \cite{wiesenfarth2021methods}. For each bootstrap sample, the final ranks were recalculated following the same evaluation protocol. We then calculated the Kendall rank correlation coefficient $\tau$ between each bootstrap-derived ranking and the original ranking computed on the full test set, and visualized the ranking variability of each team using a rank stability plot.

In the CT task, the bootstrap analysis yielded a mean Kendall’s $\tau$ of 0.9163, with a 95\% confidence interval of [0.807, 1.000], indicating a good agreement between the bootstrap-derived rankings and the original ranking. The X-ray task exhibited even higher stability, with a mean Kendall’s $\tau$ of 0.9997 and an extremely narrow 95\% confidence interval of [0.999, 1.000]. Fig.~\ref{fig:ranking} illustrates the robustness and variability of the ranking outcomes, with larger and more concentrated dots indicating greater ranking stability.

\subsection{Example Failure Cases}
\begin{figure}[t]
\includegraphics[width=\columnwidth]{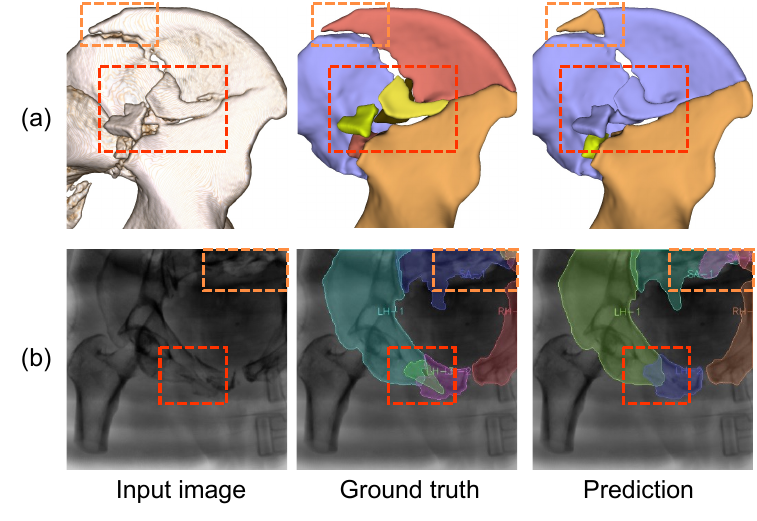}
\centering
\caption{Example failure modes. The red and yellow boxes indicate false-negative and false-positive regions, respectively.}\label{fig:failure_cases}
\end{figure}

Fig.~\ref{fig:failure_cases} presents representative failure cases from both the CT and X-ray tasks, illustrating common error patterns observed across participating algorithms.

In CT images, false negatives (missed fragments) predominantly occur in cases involving very small fragments, particularly near the sacroiliac joint, where anatomical segmentation may under-segment the surrounding bone and obscure fragment boundaries (see Sec.~\ref{FP_analysis}). Conversely, false positives (spurious fragments) often arise in incomplete or non-displaced fractures, where subtle cortical irregularities may be over-interpreted as separable fragments (see Sec.~\ref{annotation_uncertainty}). 

In X-ray images, both false negatives and false positives are primarily driven by occlusion and projection overlap. For example, superimposed anterior-ring fragments in inlet projections can render multiple fragments indistinguishable, leading to missed detections. Similarly, high-frequency edges introduced by bowel gas or soft-tissue–bone overlap may be mistakenly identified as fracture boundaries. These examples highlight the intrinsic ambiguity of 2D projections and the difficulty of recovering 3D fragment topology from overlapping structures (see Sec.~\ref{discussion_scheme_xray}).

\section{Discussion} 

\subsection{The Overall Scheme for Fragment Segmentation}
\subsubsection{CT Segmentation}
\label{discussion_scheme_ct}

In the CT task, two-stage approaches consisting of anatomical segmentation followed by fragment segmentation were widely adopted by the participants. This two-stage strategy was chosen for multiple reasons: (1) The sacroiliac joint, whether dislocated due to trauma or not, exhibits distinct characteristics compared to fracture surfaces within individual pelvic bones, making it difficult to effectively address both in a single-step segmentation approach; (2) The effectiveness of this scheme has been previously validated in the FracSegNet study, where bone region extraction simplified the task; (3) Pre-trained large models, such as TotalSegmentator and the CTPelvic1K baseline model, could be directly utilized to extract the bone regions; and (4) The evaluation metric design explicitly required accurate anatomical segmentation as a distinct criterion, ensuring alignment with the requirements of subsequent surgical planning procedures~\cite{liu2025preoperative}.
We computed the correlation between IoU-A and IoU-F in the CT task across all participating teams. A strong linear relationship was observed, with Pearson’s correlation coefficient of 0.870 (\textit{p} = 0.0005), suggesting that accurate anatomical segmentation serves as a critical foundation for fracture segmentation.

A key methodological divergence among the top-performing teams in the CT task was their approach to instance-level fracture segmentation in the second stage of the pipeline. Given the challenge of predicting a variable number of fragments, teams reformulated the problem into a semantic segmentation task using one of two main strategies: primary-secondary labeling or boundary-core decomposition. The primary-secondary strategy assigns a single dominant (primary) fragment label to each bone and groups the remaining pieces as secondary fragments, whereas the boundary-core approach segments the core regions of fragments and delineates their boundaries to enable instance separation.

Each strategy offers trade-offs. The primary-secondary method is conceptually simple and well-suited for most fracture patterns, but its effectiveness diminishes in cases where the designation of a primary fragment is ambiguous—such as bilateral fractures or when secondary fragments are of comparable size. Additionally, secondary fragments that are spatially adjacent may become fused during post-processing, reducing instance accuracy. Conversely, the boundary-core approach facilitates more precise separation of contacting or overlapping fragments, which is especially beneficial in complex fracture regions like the anterior pelvic ring. However, this method is sensitive to small boundary segmentation errors, which may disconnect or erroneously merge fragments, particularly in the presence of hairline or incomplete fractures. 

The ABBC formulation adopted by the winning MIC-DKFZ team demonstrated a clear performance advantage, particularly in complex multi-fragment cases. By explicitly modeling fragment separability, the method first predicts the core region of each fragment and then learns a thin boundary representation that delineates where fragments separate. This reduces dependence on heuristic post-processing steps and substantially lowers the risk of fragment merging that commonly arises in primary–secondary pipelines. While the boundary–core approach introduces increased sensitivity to local boundary mis-segmentation and requires careful control of splitting behavior, its strong ability to robustly separate fragments that are in close apposition appears to be a key factor behind the superior IoU and surface distance metrics observed. A broader takeaway is that explicitly modeling fracture boundaries—rather than relying solely on semantic labeling—offers a principled advantage in anatomies where structural continuity is disrupted in complex and clinically variable ways. Moreover, because the formulation does not depend on bone shape priors, it can be directly applied to other fracture anatomies without requiring changes to the labeling policy.

\subsubsection{X-Ray Segmentation}
\label{discussion_scheme_xray}

In contrast to the relatively consistent two-stage pipelines in the CT task, the X-ray task showcased greater methodological diversity, particularly in the overall segmentation schemes adopted by participating teams. These differences reflect the inherent challenges of 2D projection imaging, where occlusion and overlapping structures complicate the delineation of fracture fragments.

As shown in Fig.~\ref{fig:Fig_scheme_xray} and Table~\ref{tab:algorithm_xray}, the most notable distinction among top-performing teams lies in how they structured their overall approach. Some teams, such as SMILE and LME, employed a two-stage strategy—first isolating the pelvic bones, followed by fragment segmentation. This separation can simplify the learning problem for the fragment segmentation module by removing irrelevant background and anatomical clutter. However, it may introduce compounding errors: inaccuracies in the anatomical segmentation stage could propagate and affect downstream fragment predictions.

Other teams, like MedIG and luckyjing, opted for a single-stage strategy that directly outputs either semantic or instance-level labels from the original input image. Notably, MedIG used three independent networks for segmenting the three bone regions separately, whereas others used a single network with multi-label prediction. Using separate networks allows specialization and potentially better accuracy for each bone, especially when the visual characteristics or fragment patterns differ, but at the cost of computational efficiency. In contrast, using a single network with shared backbone and multi-class output offers better efficiency and enforces global anatomical coherence, but may suffer from performance degradation in locations that are underrepresented or visually ambiguous.

Efficiency considerations are especially relevant for the X-ray task, where segmentation may be required in time-sensitive clinical settings such as emergency trauma care or intra-operative fluoroscopic guidance. While we report container runtimes as an efficiency proxy, these measurements include substantial overhead from environment initialization and I/O, and therefore do not reflect pure model inference time. Additionally, several teams employed model ensembling and repeated inference strategies to maximize accuracy within the runtime limit, trading computational efficiency for performance. Notably, the Mask R-CNN-based pipelines demonstrated faster inference on 2D images and, with appropriate pruning or lightweight backbone substitution, may be adaptable for near real-time use. In contrast, two-stage anatomical-then-fragment pipelines, though accurate, incur higher computational load and are less compatible with fast-turnaround workflows.

Across all methods, the target remained consistent: fragment-level labeling, rather than fracture-line segmentation, as the latter is confounded by projection overlap in X-rays. Teams that employed the primary-secondary labeling, such as SMILE, faced challenges in separating small fragments, which tended to appear fused in the projection space. While instance segmentation models like Mask R-CNN (LME and luckyjing) offer potential for explicit fragment separation, they remain limited by the lack of depth cues and can struggle with over-segmentation or missing small fragments.

These varying design choices underline the difficulty of the X-ray segmentation task and point to important trade-offs between modularity, anatomical specificity, and model generalization. 
Despite the similar task formulation, the segmentation performance on X-ray images was notably lower than on CT. For instance, the top-performing method in the CT task achieved a IoU-F of 0.93, whereas the top method in the X-ray task only reached 0.77. From a clinical perspective, such a discrepancy has substantial implications: in surgical planning or intraoperative guidance, insufficiently separated or misidentified fragments may lead to incorrect assessments of fracture morphology or fragment displacement, ultimately affecting surgical strategy.

The reduced accuracy in X-ray segmentation arises from intrinsic limitations of projection imaging: (1) A single 3D fragment may appear disconnected or distorted in 2D projections due to occlusion or truncation by the limited field of view; and (2) Overlapping anatomical structures frequently obscure fracture lines, especially in complex or comminuted injuries, making instance-level identification extremely challenging. These factors collectively compromise the ability of current semantic and instance segmentation methods to distinguish fragments robustly in X-rays.

To bridge this performance and applicability gap, future efforts could incorporate prior anatomical knowledge or leverage additional imaging information. One strategy is the acquisition of multiple views or targeted X-ray angles; complementary perspectives could help disambiguate fragment boundaries that are occluded in a single view. Another direction is to integrate preoperative CT segmentation via 2D/3D registration~\cite{grupp2019pose}, providing a rough anatomical prior for localizing fragments in intraoperative X-rays. However, practical deployment of this approach faces challenges, including estimation of C-arm pose and compensating for changes in patient positioning or fragment displacement between scans.

Although these strategies exceed the scope of the current challenge, they reflect the necessary steps toward translating X-ray segmentation into reliable clinical tools—particularly in image-guided orthopedic procedures, where real-time, accurate fragment localization is critical.

\subsection{Impact of Segmentation on Reduction Planning}
\begin{figure}[t]
\includegraphics[width=\columnwidth]{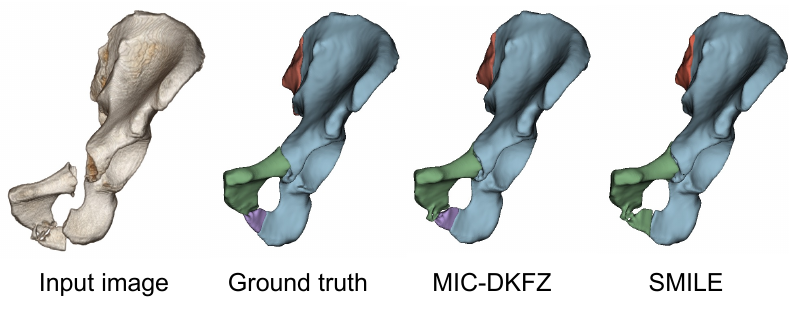}
\centering
\caption{Example reduction planning result using different segmentation approaches. The anatomical alignment is affected by the separability of fragments.}\label{fig:reduction_planning}
\end{figure}

To demonstrate the clinical implications of segmentation quality, particularly its impact on downstream surgical planning, we conducted a comparative analysis using the FracFormer reduction planning framework~\cite{yibulayimu2025fracformer}. We evaluated reduction outcomes derived from two representative CT segmentation strategies: the boundary–core decomposition used by MIC-DKFZ and the primary–secondary labeling scheme used by SMILE.

As illustrated in Fig.~\ref{fig:reduction_planning}, differences in fragment separability directly propagate into the reduction stage. When small fragments are not correctly isolated—an issue more commonly observed in primary–secondary pipelines—fused or misidentified fragments introduce incorrect geometric constraints during alignment, causing the estimated pose to drift away from the desired anatomical configuration. This observation is consistent with prior findings showing that manually separating small fragments on automated segmentation results can improve reduction pose estimation~\cite{liu2025preoperative}. In the example shown, the boundary–core segmentation produced a chamfer distance of 1.695 mm, compared with 2.097 mm for the primary–secondary result. Although both errors fall within clinically acceptable ranges, this experiment underscores a key insight: instance-level separability—rather than volumetric overlap alone—is critical for downstream tasks that depend on accurate geometric reconstruction, such as reduction pose/trajectory estimation and fixation strategy design.

\subsection{Limitations and Future Directions}
\subsubsection{Dataset}

Compared to related challenges organized at major medical imaging conferences, where the number of CT samples typically ranges from 160 to 1,400, the PENGWIN dataset is relatively small in size. This limitation is primarily due to the rarity of pelvic fracture CT data, as pelvic fractures account for only 3\% of all fractures~\cite{hu2023epidemiology}. Additionally, all cases were collected from clinical centers in China, representing a relatively homogeneous patient population. This may limit the generalizability of trained models across different demographic groups and imaging practices.

While the X-ray dataset closely resembles real clinical data due to the DeepDRR simulation approach, it remains synthetic rather than derived from real patient imaging. Although DeepDRR facilitates high-fidelity X-ray generation, enabling this prospective study with a large annotated dataset, the absence of real C-arm X-ray images raises concerns regarding its direct clinical translation. Due to the 2D projection nature, inter-observer variability was not evaluated in the X-ray task, as consistent delineation of overlapping fragments was not achievable even among trained annotators. 
In addition, multiple X-ray projections were generated from each CT scan, and individual projections were treated as independent test samples during evaluation. While variation in projection angle leads to meaningful differences in occlusion and visibility patterns, multiple projections per subject introduce repeated measures and reduce strict statistical independence in the test set. 

For future iterations of PENGWIN, we plan to substantially expand the fractured CT dataset, increasing the number of cases to 300–500 scans. Furthermore, we aim to incorporate real intraoperative C-arm X-ray images into the test set. Since fragment-level annotation in X-ray images is inherently challenging, these images will be labeled using a 2D/3D registration pipeline to align them with preoperative CT segmentations, ensuring anatomical consistency and clinically meaningful ground truth. This will also allow us to evaluate sim-to-real generalization performance, building on recent findings showing that DeepDRR-generated synthetic data can support transfer to real clinical X-rays~\cite{gao2023synthetic}.

\subsubsection{Inaccuracy in Annotation}
\label{annotation_uncertainty}

The initial anatomical segmentation used to support the annotation workflow was generated by a trained automatic model and subsequently refined by expert annotators. While this approach significantly reduced manual workload and improved structural consistency, it may introduce subtle bias from the model’s prior assumptions.

As discussed in Sec.~\ref{inter-observer}, the refinement process involved inherent uncertainty, requiring annotators to determine whether an incomplete fracture should be classified as two separate fragments or a single piece. This decision was primarily based on their understanding of the subsequent surgical process, particularly whether the two segments could be manipulated as a single unit during reduction without undergoing significant deformation.

In clinical practice, this determination is made comprehensively by the surgeon, considering not only medical imaging but also physical examinations and additional clinical information, such as the mechanism of injury. Given this complexity, a collaborative interaction between human experts and segmentation algorithms may be necessary. The current approaches, particularly the primary-secondary labeling scheme, still require manual refinement by operators, which may involve merging or splitting labels. This process remains time-consuming, especially when a larger number of fracture surfaces require delineation.

For future iterations of PENGWIN, we aim to explore interactive fragment segmentation. One potential direction is to incorporate a fast manual seed point selection step, allowing operators to define the number and approximate positions of fragments before segmentation. This approach is expected to eliminate uncertainty regarding fragment count while enhancing segmentation performance by incorporating prior anatomical knowledge into the process.

Another potential direction is to integrate uncertainty quantification, allowing the network to generate multiple segmentation results that reflect different interpretations of incomplete fractures or unclear fracture lines. This approach would enable operators to select the most appropriate segmentation scheme, with their attention guided toward highlighted regions of uncertainty for further assessment.

\subsection{Recommendations for Future Challenges}

We summarize several practical lessons identified during the challenge and the post-challenge participant survey, which should be addressed in future iterations and may serve as useful guidance for other challenge organizers.

First, although the dataset and preliminary validation phase were released well in advance, the final test phase deadline was extended by three days due to multiple participant requests. To maintain fairness, teams that had already submitted were given an opportunity to update their submissions. However, such extensions should be avoided in future iterations to uphold the rigor and integrity of the challenge timeline.

Second, apart from providing an algorithm template, we did not offer a baseline method for participants to build upon, which may have contributed to delays in submissions. Including a reference baseline in future iterations could help standardize development efforts and provide a clearer starting point for participants.

Lastly, we encountered challenges in retrospectively collecting algorithm descriptions from participants after the challenge had ended, particularly from lower-performing teams. To address this, future challenges may require participants to submit a technical description of their method along with their final submission, ensuring that essential technical details are documented in a timely manner.

\section{Conclusion} 

We have presented the PENGWIN challenge on pelvic fracture segmentation in CT and X-ray images, organized as part of MICCAI 2024. We provided a comprehensive review of related challenges and studies, a detailed overview of the challenge setup, dataset, participant contributions, and performance analysis. 

In the CT task, the winning team achieved a high level of accuracy, with an average fragment-wise IoU of 0.930. 
In the X-ray task, although the top-performing method reached an IoU of 0.774, this level of accuracy remains insufficient for reliable clinical use. Future work will incorporate evaluation on real C-arm X-ray datasets and explore multi-view or CT-informed approaches to mitigate projection ambiguity.

Beyond quantitative performance, the participating algorithms demonstrated great methodological diversity, offering valuable insights for future research in this emerging field. The various approaches and strategies employed have been thoroughly discussed in this paper. While the focus of this challenge was on pelvic fractures, we believe that many of the developed methodologies and key findings will generalize to other types of fractures.

\bibliographystyle{IEEEtran} 
\bibliography{refs_PENGWIN.bib}

\end{document}